\documentclass[11pt,a4paper]{article}
\pdfoutput=1
\usepackage{jheppub}
\usepackage{rotating}
\usepackage{array}
\usepackage{amsmath}
\usepackage[normalem]{ulem}
\usepackage{slashed}
\usepackage{booktabs}
\usepackage[pdftex,table,dvipsnames]{xcolor}
\usepackage{units}
\usepackage{multirow}
\usepackage{url}
\usepackage{parskip}
\usepackage[all]{nowidow}
\usepackage{placeins}
\usepackage{cleveref}

\usepackage{xspace}
\usepackage{subfig}
\usepackage{color}

\usepackage{dsfont}
\usepackage{soul}
\usepackage{hyperref}

\usepackage{bm} 

\allowdisplaybreaks

\newcommand{\be}{\begin{equation}}
\newcommand{\ee}{\end{equation}}
\renewcommand\({\left(}
\renewcommand\){\right)}
\renewcommand\[{\left[}
\renewcommand\]{\right]}
\newcommand{\dd}{{\rm d}}

\newcommand\eps{\epsilon}

\newcommand\GeV{{\rm GeV}}

\let\vec\mathbf

\newcommand\nn{\nonumber}

\def\O{\mathcal{O}}

\def\R{\mathcal{R}}
\def\P{\mathcal{P}}

\preprint{Nikhef 2025-001}

\title{Scattering meets absorption in dark matter detection}

\author[1,2]{Pieter~Braat,}
\author[1,3]{Anh Vu~Phan,}
\author[1,3]{Marieke~Postma,}
\author[1,3]{and Susanne~Westhoff\,}

\affiliation[1]{Nikhef, Science Park 105, 1098 XG Amsterdam, The Netherlands}
\affiliation[2]{Institute of Physics, University of Amsterdam, Science Park 904, 1098 XH Amsterdam, The Netherlands}
\affiliation[3]{Institute for Mathematics, Astrophysics and Particle Physics, Radboud University, 6500 GL \mbox{Nijmegen}, The Netherlands\\}

\emailAdd{pbraat@nikhef.nl}
\emailAdd{anhvu.phan@ru.nl}
\emailAdd{mpostma@nikhef.nl}
\emailAdd{susanne.westhoff@ru.nl}

\abstract{Direct detection experiments have started to explore dark matter scattering off electrons and nucleons through light mediators. Mediators with sub-keV  masses are efficiently produced in the Sun and can be absorbed in the same detectors that probe dark matter scattering. We investigate the interplay of dark matter scattering and mediator absorption for two models with a dark photon as mediator. For Dirac dark matter, we find that scattering and absorption can be simultaneously observed at direct detection experiments in the near future. For atomic dark matter, we predict additional signals due to scattering of both dark atoms and constituents from ionized dark atoms. In both models, we determine the parameter space that respects bounds from cosmology and astrophysics, where the strongest constraints come from  dark matter self-interactions.  In this way, we identify viable targets for dark matter with light mediators at upcoming direct detection experiments. Distinguishing between the various signals, for instance by measuring energy distributions, will be crucial to reveal the underlying model in case of a discovery.}

\begin{document}

\maketitle

\flushbottom

\section{Introduction}
\label{SEC:introduction}
Current direct detection experiments searching for dark matter-electron or dark matter-nucleon scattering have unprecedented sensitivity to dark matter (DM) interactions with electrons or nucleons~\cite{SENSEI:2023zdf,DarkSide:2022dhx,DarkSide:2022knj,XENON:2023cxc,XENON:2019zpr,XENON:2019gfn,PandaX:2023xgl,PandaX-II:2018xpz}. In particular, the advent of sub-GeV dark matter searches has shown that scattering through light force mediators entails a high sensitivity to the underlying couplings \cite{Battaglieri:2017aum,Essig:2011nj,Knapen:2017xzo}. In addition to scattering, direct detection experiments can probe the absorption of bosonic matter from the halo or produced in the Sun~\cite{An:2013yua,Bloch:2016sjj,Hochberg:2016sqx,An:2020bxd}. These two types of signals have been searched for independently, and the results have been interpreted in terms of separate models of dark matter.

We investigate for the first time if dark matter scattering and the absorption of mediators emitted from the Sun can occur \emph{simultaneously} in a given model. For two models with (sub)-keV-scale vector mediators, we show that simultaneous scattering and absorption signals are a natural prediction with important consequences for the interpretation of direct detection searches.

A very sensitive astrophysical probe of light force mediators is stellar cooling~\cite{An:2013yfc,Redondo:2013lna,Hardy:2016kme}. Luminosity measurements of stars like the Sun or red giants strongly constrain the stellar production and emission of bosons with masses from a few meV up to several hundreds of keV~\cite{An:2013yfc,Chang:2016ntp}. For scalar mediators, stellar cooling bounds are stronger than the sensitivity to absorption at current direct detection experiments. For vector mediators, bounds from stellar cooling decouple in the massless limit due to in-medium effects~\cite{An:2013yfc}. In particular, current direct detection experiments are sensitive to the absorption of kinetically mixed dark photons from the Sun, despite the strong bounds on their emission from stellar cooling~\cite{An:2020bxd}. We therefore focus on models with dark photons with meV-to-keV masses as mediators.

Light force mediators generally lead to sizeable dark matter self-interactions, which tend to have drastic effects on structure formation in the early universe. At high relative velocities, dark matter self-interactions are constrained by the Bullet Cluster observation, see e.g.~\cite{Randall:2008ppe,Markevitch:2003at}; at lower velocities, self-interactions could address apparent discrepancies between observations and predictions based on collisionless cold dark matter~\cite{Moore:1994yx,Flores:1994gz,Walker_2011,deBlok:2001hbg,deBlok:2002vgq,Simon:2004sr,Tulin:2017ara,Adhikari:2022sbh}. We take these constraints into account when we make predictions for direct detection experiments.

For concreteness, we consider two specific models for dark matter: Dirac Dark Matter~\cite{Pospelov:2007mp,Knapen:2017xzo}, where the dark matter candidate is the lightest fermion in a QED-like dark sector; and Atomic Dark Matter~\cite{Kaplan:2009de,Kaplan:2011yj}, where dark matter consists of bound states of dark fermions, which can be partly ionized in the relic abundance today. In both models, the mediator to the Standard Model is a kinetically mixed massive dark photon. For Dirac Dark Matter, we analyze the interplay of dark fermion scattering with nucleons and electrons and dark photon absorption at direct detection experiments. For Atomic Dark Matter, we investigate the intricate relations between dark atom and dark constituent scattering and combine them with dark photon absorption. Our predictions set concrete targets for dark matter searches with light vector mediators at upcoming direct detection experiments.

This article is structured as follows. In \cref{SEC:models}, we introduce the two dark matter models. In \cref{SEC:general-constraints}, we review and analyze constraints on these models from cosmology and astrophysics and introduce the formalism for direct detection searches. For the cosmologically viable scenarios of Dirac Dark Matter and Atomic Dark Matter, we present our predictions of signals at direct detection experiments in \cref{SEC:Dirac_DM,SEC:atomic-dm}, respectively. We conclude in \cref{SEC:conclusions}. Details on our calculation of dark atom-nucleon scattering can be found in \cref{sec:app1}.

\section{Dark matter models}
\label{SEC:models}
We consider a dark sector with a $U(1)_d$ gauge symmetry and Dirac fermions $\chi$, which are charged under $U(1)_d$. The symmetry is broken by either a dark Higgs field or the St\"uckelberg mechanism, providing a mass $m_d$ for the gauge field $A_d$, the dark photon. The dark sector and the Standard Model couple through kinetic mixing \cite{Holdom:1985ag} of the dark gauge boson with the hypercharge field $\mathcal{L} = - \frac12\, \bar \eps\, F_{d}^{\mu\nu} B_{\mu\nu}$, where $F_{d}^{\mu\nu},\,B^{\mu\nu}$ are the $U(1)_d$ and hypercharge field strength tensors and $\bar \eps$ is the mixing parameter. After electroweak symmetry breaking the dark photon mixes with both the photon and the $Z$ boson, with mixing parameters $\eps \equiv \bar \eps \cos \theta_W$ and $-\bar \eps \sin \theta_W$, respectively, where $\theta_W$ is the weak mixing angle. Here we focus on mixing with the QED sector, which dominates the low-energy phenomenology.

The kinetic mixing can be `rotated away' by redefining the photon field $A^\mu \mapsto A^\mu + \eps A_d^\mu$, which keeps the mass matrix diagonal. The relevant part of the Lagrangian becomes
\begin{align}\label{eq:lagrangian}
    \mathcal{L} \supset \sum_i \overline{\chi}_i(i\gamma^\mu\partial_\mu-m_{\chi_i}) \chi_i  +\frac{m_d^2}{2} A_{d}^{\mu} A_{d,\mu} - (g_d\, j^\mu_d -\eps e j^\mu_{\rm em} ) A_{d,\mu} - e\, j^\mu_{\rm em} A_{\mu}\,,
\end{align}
with the sum running over the dark Dirac fermions with masses $m_{\chi_i}$. The electromagnetic and dark gauge couplings are denoted by $e$ and $g_d$ respectively. The last two terms in the Lagrangian describe the fermion - (dark) photon couplings. The SM fermions pick up a millicharge under the dark gauge group $U(1)_d$. The dark and SM electromagnetic fermion currents are
\be
j^\mu_d  = \sum_i q_i\, \overline{\chi}_i \gamma^\mu \chi_i\,, \qquad
j^\mu_{\rm em}= \sum_\alpha Q_\alpha \bar{f}_\alpha \gamma^\mu f_\alpha\,,
\ee
with $q_i$ the charges of the dark-sector fermions. The sum over $\alpha$ runs over all SM fermions $f_\alpha$ with electric charge $Q_\alpha$. 

The dark photon mass $m_d$ breaks the $U(1)_d$ symmetry. We will assume that $m_d$ is generated by the St\"uckelberg mechanism~\cite{Feldman:2007wj}. If a dark Higgs mechanism is at the origin of the dark photon mass~\cite{Gopalakrishna:2008dv}, the dark Higgs field can have a significant impact on the phenomenology of the dark matter models~\cite{Ahlers:2008qc,An:2013yua}. We comment on this alternative scenario wherever it is relevant. 

We consider two models for dark matter:
\begin{enumerate}
\item Dirac Dark Matter~\cite{Pospelov:2007mp,Knapen:2017xzo}: The lightest fermion in the dark sector, the dark electron $\chi= e'$,  is the dark matter candidate. We assume that all additional dark fermions, if present in the model, are sufficiently heavy that they do not affect the low-energy phenomenology discussed in this work. Without loss of generality, we set $q_{e'}=-1$. 
\item Atomic Dark Matter \cite{Kaplan:2009de,Kaplan:2011yj}: The dark sector contains two dark fermions $\chi=\{e',p'\}$. The dark matter candidate is a bound state (``dark hydrogen'' $H'$) of the dark electron $e'$ and the dark proton $p'$. Without loss of generality, we take $m_{p'} \geq m_{e'}$ and unit charges $q_{e'}=-1$ and $q_{p'}=1$.
Bound-state formation requires the effective Yukawa potential from dark photon exchange to be sufficiently long-ranged. For $s$-wave bound states, the mass of the dark photon must be smaller than the Bohr momentum of the system~\cite{PhysRev.125.1131}
\begin{align}
    m_d < 1.2\,\alpha_d\, \mu_{H'}\,,\qquad \mu_{H'} = \frac{m_{e'} m_{p'}}{m_{e'} + m_{p'}}= \frac{ m_{p'}}{1+R}\,,
    \label{bohr}
\end{align}
with $\mu_{H'}$ the reduced mass of the dark hydrogen bound state, and $\alpha_d = g_d^2/4\pi$ the dark fine-structure constant. We require the gauge coupling to be perturbative, $\alpha_d \lesssim 1$. This also ensures that the binding energy $B_{H'} = \frac12\alpha_d^2\mu_{H'}$ is smaller than the sum of the constituent masses. For future reference, it is useful to define the mass ratios \cite{Cline:2013pca}
\begin{align}
  R \equiv \frac{m_{p'}}{m_{e'}}\,, \qquad
   f(R) \equiv \frac{m_{H'}}{\mu_{H'}} = R + 2 + \frac{1}R -\frac12 \alpha_d^2\,.
    \label{R}
\end{align}
We will neglect the coupling-dependent contribution from the binding energy, the last term in $f(R)$.

\end{enumerate}

\section{General constraints on Dirac/Atomic Dark Matter}
\label{SEC:general-constraints}
Both Dirac Dark Matter and Atomic Dark Matter are constrained by cosmology, astrophysics and searches at direct detection experiments, see e.g.~\cite{Knapen:2017xzo,Bernal:2019uqr,Buen-Abad:2021mvc,Cline:2021itd}, which together limit the viable parameter space considerably. In this section, we summarize the existing constraints.

We are interested in scenarios where both dark matter scattering and dark photon absorption can be observed. This restricts the viable parameter space to feebly coupled dark sectors with kinetic mixing below $\eps \lesssim 10^{-8}$. Such small mixing is beyond the sensitivity of current collider and beam-dump experiments (see Ref.~\cite{Gori:2022vri} for a recent summary of the constraints). Small mixing also prevents dark matter annihilation from causing observable effects in the cosmic microwave background (CMB) or in indirect detection experiments~\cite{Cirelli:2024ssz}.\footnote{Indirect searches for dark matter annihilation through trident dark photon decays~\cite{Linden:2024uph} are capable of probing small kinetic mixing for dark photon masses between 100\,keV and 1\,MeV. However, this parameter range falls beyond the range of interest in this work.}
The mass range of interest in our analysis is $m_{\rm DM} = {\rm MeV} \, -\, {\rm TeV}$ for dark matter, and $m_d = {\rm meV}\,-\, {\rm eV}$ for dark-photon mediators, such that they fall within the sensitivity range of direct detection experiments. Depending on the model, $m_{\rm DM}$ stands for either the dark electron mass or the dark hydrogen mass.

\subsection{Cosmic Microwave Background and nucleosynthesis}\label{sec:cmb}

The Planck data on the cosmic microwave background~\cite{Planck:2018vyg} limits the amount of dark radiation and its coupling with dark matter. The constraints only apply to dark photons that are relativistic at the time of recombination $T_{\rm rec} \approx 0.26\,$eV. For larger dark photon masses the formation of light elements during big bang nucleosynthesis (BBN)  at temperatures $T_{\rm BBN} \sim \,$MeV still limits the dark radiation density.

The CMB bound on the amount of dark radiation depends on the light degrees of freedom in the dark sector at recombination, which bounds the dark-sector temperature $T_d$~\cite{Cline:2021itd} 
\begin{align}\label{eq:cmb}
    \xi = \frac{T_d}{T} \Big|_{z=0} < 0.52 \( \frac{3}{h_{d}(T_{\rm rec})}\)^{1/4},
\end{align}
with  $T$ the temperature of the SM radiation. If only the dark photon is relativistic at recombination, the light degrees of freedom in the dark sector are $h_{d}(T_{\rm rec})=3$.\footnote{The massive St\"uckelberg photon has three degrees of freedom. If the dark photon mass is generated by a dark Higgs mechanism, the  high-temperature degrees of freedom are the two transverse photon polarizations and the two degrees of freedom of the complex Higgs boson.}
This bound is slightly stronger than the limit from Big Bang Nucleosynthesis (BBN)~\cite{Cyr-Racine:2012tfp}, although the latter is valid at higher temperatures $T_{\rm BBN} \sim 1\,$MeV, when potentially more dark-sector degrees of freedom are light. 

If the dark and visible sectors decouple kinetically (shortly) before the QCD phase transition $T_\text{QCD}\sim 150\,\text{MeV}$, the temperature ratio today is 
\begin{equation}\label{eq:xi}
   \xi \sim 0.3 \(\frac{h_d(T_\text{QCD})}{h_d(T_\text{rec})}\)^{1/3}.
\end{equation}
This translates to $\xi \sim 0.3-0.5$, depending on the masses of the dark sector particles. For example, if the dark electron, proton and photon are all lighter than 150\,MeV, then $h_d(T_{\rm QCD}) =11$ and assuming again that only the dark photon is relativistic at low temperatures $h_d(T_{\rm rec}) =3$ gives   $\xi \sim 0.46$, if no additional heating mechanisms are present in either sector. The CMB bound~\eqref{eq:cmb} is then automatically satisfied. In the parameter space of interest, however, the dark sector fermions are always significantly heavier than 150\, MeV, in which case $\xi = 0.3$. Alternatively, if the two sectors were never in thermal contact - for instance, in Dirac dark matter models where the relic abundance is set via a freeze-in mechanism - the initial temperature ratio is a free parameter of the model.
We will use  $\xi = 0.3$ for our numerical results; the only (relevant) quantity that depends on this choice is the ionization fraction \cref{ion} in the Atomic Dark Matter model. 

If the dark sector contains dark radiation that is strongly coupled to (a fraction of) the dark matter Dark Acoustic Oscillations (DAO) may occur, similarly to the Baryon Acoustic Oscillations in the visible sector \cite{Cyr-Racine:2012tfp,Cyr-Racine:2013fsa}. The strong coupling prohibits the dark matter  from clumping until it decouples from the dark radiation bath. Effectively, the DAO generates a characteristic length scale $r_{\rm{DAO}}$, below which structure formation is delayed and which is imprinted in the matter power spectrum: at scales below $r_{\rm{DAO}}$ the matter power spectrum is suppressed. The DAO scale 
is set by the parameter $\Sigma_{\rm{DAO}}$, implicitly defined via  the dark matter - dark radiation interaction cross section \cite{Cyr-Racine:2013fsa}
\be
\( \frac{\sigma_{\rm DM-DR}}{m_{\rm DM}}\) = 6 \times 10^{-6} \( \frac{\xi}{0.5}\) \(\frac{\Sigma_{\rm DAO}}{10^{-4.5}}\) \,\frac{{\rm cm}^2}{\rm g}.
\label{Sigma_gen}
\ee
The CMB data imposes the bound \cite{Cyr-Racine:2013fsa} 
\begin{equation}
    \Sigma_{\mathrm{DAO}} \lesssim 3\times 10^{-5}\,,
    \label{Sigma_bound}
\end{equation}
assuming that all dark matter couples to the dark radiation. For larger $\Sigma_\text{DAO}$, the allowed fraction of dark matter quickly goes to zero.

More recently, the DAO were studied in \cite{Bansal:2022qbi}, where a bound on $r_\text{DAO}$ is set that is roughly $40\%$ more stringent than earlier works. This translates into a slightly stronger bound on $\Sigma_\text{DAO}$ as well. As the DAO constraint does not significantly impact the parameter space for both Dirac and Atomic Dark Matter, we use \cref{Sigma_bound} to estimate its effect.

\subsection{Bullet Cluster and small-scale structure}

Dark matter self-interactions mediated by dark photon exchange can affect structure formation on small scales.
The Bullet Cluster observation sets an upper bound on the transfer cross section $\sigma_T =\int \dd \Omega (1-\cos \theta) \frac{\dd \sigma}{\dd \Omega}$ for self-interactions~\cite{Randall:2008ppe,Markevitch:2003at}
\footnote{The conversion factors to natural units are 
$ {\rm MeV}^{-2} = (1.97 \times10^{-11} {\rm cm})^2$ and
${\rm MeV}^{-3} =2.18\times10^5\, \mathrm{cm^2/g}$.}
\be
{\rm Bullet} \;\;{\rm Cluster}: \quad \frac{\sigma_T}{m_{\rm DM}} \lesssim 1\,{\rm cm^2/g} \qquad {\rm for} \; v= 1000\,{\rm km/s},
\label{bullet}
\ee
where $v$ is the relative velocity of the dark matter inside the colliding clusters.

Small-scale structure observations are sensitive to dark matter self-interactions at smaller velocities
$v \sim 10\,{\rm km/s}$. However, the resulting bounds are affected by large uncertainties both in structure formation simulations and in observations (see the review \cite{Adhikari:2022sbh})
\be
{\rm small} \;\;{\rm scales}: \quad \frac{\sigma_T}{m_{\rm DM}} \lesssim c^{\rm ss}\,{\rm cm^2/g} \qquad {\rm for} \; v= 10\,{\rm km/s},
\label{small_scale}
\ee
where values $c^{\rm ss}= 0.1 -100$ are quoted in the literature~\cite{Zavala:2019sjk,Correa:2021qam,Silverman:2022bhs}. 

If the dark matter self-interactions  are velocity-dependent, they can have  sizeable effects in small-scale structures, while still satisfying the Bullet Cluster bound. This opens up the possibility to explain apparent discrepancies between observations of small-scale structures and simulations  based on cold dark matter \cite{Navarro:1996gj,Navarro:1995iw,Moore:1994yx,Flores:1994gz,Walker:2011zu,deBlok:2001hbg,deBlok:2002vgq,Simon:2004sr} 
(see e.g. \cite{DelPopolo:2021bom} for a recent review) through dark matter self-interactions \cite{Spergel:1999mh,Kaplinghat:2015aga}.

In our analysis, we will consider the Bullet Cluster bound \cref{bullet} as a hard constraint, while for small scales \cref{small_scale} we show results for the range $c^{\rm ss} = 1-100$. Values around $c^{\rm ss} = 1$ can be viewed as an indication of interesting small-scale dynamics, whereas values close to $c^{\rm ss} = 100$ should be considered as a bound.

If the fraction of self-interacting dark matter $f$ is less than about $30\%$, it is too small to  have an observable impact on the Bullet Cluster dynamics and the constraint \cref{bullet} disappears~\cite{Markevitch:2003at,Feng:2009mn}. Likewise, we expect observable effects on small-scale structures to disappear for $f< 1-10\%$ \cite{Boddy:2016bbu,Knapen:2017xzo}.

\subsection{Stellar cooling}

Dark photons emitted from horizontal branch stars or red giants cause stellar cooling, but cannot be observed in direct detection experiments, since the sources are  far away. In this work, we therefore focus on dark photons with masses in the range $m_d = {\rm meV}\,-\, {\rm eV}$, for which the Sun is an efficient source and which at the same can be absorbed with an observable rate in direct detection experiments.

The kinetic mixing in stars can differ significantly from vacuum mixing because of medium effects. In particular, if the dark photon mass is smaller than the plasma mass $m_T$ for the photon -- a few keV in the Sun, a few $10$ keV in red giants, and around 1 MeV in supernovas -- 
stellar production and absorption of dark photons is suppressed \cite{An:2013yua,An:2013yfc,An:2014twa,Hardy:2016kme,Chang:2016ntp}. 
For sub-keV dark photon masses, the strongest bound on the kinetic mixing parameter is set by absorption of solar dark photons in direct detection experiments~\cite{An:2020bxd}, discussed in \cref{sec:direct_detection}, which supersedes solar cooling constraints on the dark photon. 

Whereas the dark photon decouples in the stellar environment in the massless limit, all matter particles charged under the dark gauge group take the opposite route; they now  couple to electromagnetism. The dark matter fermions and, if present, the dark Higgs  become effectively millicharged particles with charge $|q_i^{\rm em}| = (q_i g_d/e) \eps$ \cite{Ahlers:2008qc,An:2014twa,An:2020bxd}, and the stellar cooling bounds on millicharged matter apply
\cite{Davidson:2000hf,Ahlers:2008qc,Vogel:2013raa}:
\be
|q^{\rm em}_i|  
\lesssim \left\{ 
\begin{array}{rl}
2\times 10^{-14} & \quad {\rm for} \; m_\chi \lesssim 5\, {\rm keV} \\
10^{-8} & \quad {\rm for} \; m_\chi \sim 5\,{\rm keV} - {\rm MeV}.
\end{array} \right.
\label{stellar_cooling}
\ee
For sub-keV dark fermion/dark Higgs masses, the bounds are dominated by the Sun and red giants, and for keV-MeV masses by supernovas. There are no stellar cooling bounds for larger masses.

If the dark photon mass is generated by a St\"uckelberg mechanism, the dark fermions are the only millicharged particles, and for masses $m_\chi \gtrsim \,$MeV stellar cooling bounds are absent.
If instead the dark gauge symmetry is broken by a dark Higgs mechanism, the bounds  are generically quite strong although model dependent \cite{An:2013yua,Ahlers:2008qc}, as the dark Higgs interactions responsible for cooling can be suppressed by large dark Higgs masses and/or small gauge couplings. Assuming that the kinetic mixing derives from integrating out loop effects of a heavy degree of freedom charged under both the dark and visible gauge groups fixes the freedom in the the model; the mixing parameter is then $\eps \sim g_d e/(4\pi)^2$ leading to a strong bound from stellar cooling~\cite{Cline:2024wja}
\be
\eps \lesssim \sqrt{\frac23} \frac{e m_d}{4\pi^{3/2}T} \qquad {\rm for}
\;\; \eps \sim g_d e/(4\pi)^2\label{higgs_bound}
\ee
with $T \sim m_T$ the typical scale of the processes inside the star.

\subsection{Direct detection}
\label{sec:direct_detection}

Dark matter can interact in different ways with the detector material in direct detection experiments. For Dirac Dark Matter with masses $m_{\rm DM} \gtrsim $ few GeV, experiments searching for scattering off nucleons have the highest sensitivity, while for lighter dark matter searches for electron interactions dominate.
In Atomic Dark Matter, a dark matter signal in an experiment can stem from different sources: depending on the fraction of ionized dark atoms, dark fermion-electron scattering, dark fermion-nucleon, and dark atom-nucleon scattering can all induce a potentially observable signal. 

In both models, the exchange of light dark photon induces spin-independent dark fermion-nucleon scattering and dark fermion-electron scattering. On top of this, kinetic mixing generates $Z$-boson-mediated dark fermion scattering with electrons, protons and neutrons, since the $U(1)_d$ field mixes with the hypercharge field~\cite{Bauer:2018onh}. However, these $Z$-mediated contributions 
are suppressed as $q^2/m_Z^2 \ll 1$, with $q$ the momentum exchange in the scattering and $m_Z$ the $Z$ boson mass. We neglect them in our analysis.

Besides dark matter scattering, both models predict another direct detection signal from the absorption of dark photon mediators. For masses $m_d \lesssim 1\,$keV, dark photons can be efficiently produced in the Sun and reach the Earth before they decay~\cite{An:2013yua,Redondo:2013lna}; heavier dark photon emission is strongly constrained by stellar cooling~\cite{Dolan:2023cjs}. In direct detection experiments, they are absorbed in the detector material by ionizing atoms, which produces a signal from the ejected electron~\cite{An:2013yua,Bloch:2016sjj,An:2020bxd}.

In searches for dark matter-electron scattering, which rely on ionization probed through an `S2-only' signal at XENON1T \cite{XENON:2019zpr} and PandaX \cite{PandaX:2023xgl}, dark photon absorption can add on top of dark matter scattering. In presence of a signal, distinguishing between absorption and scattering would be possible by analyzing the energy spectrum of the signal, if statistically possible. In current searches, simultaneous absorption and scattering is relevant for dark matter with masses between about $1$ and $6\,$GeV. Signals from lighter dark matter candidates fall below the energy detection threshold, while for heavier candidates searches for nucleon scattering through a combined S1-S2 signal in PandaX \cite{PandaX:2023xgl,PandaX-II:2018xpz} are more sensitive than an S2-only analysis.

Investigating the interplay between different signals in a single experiment, and between individual signals in different experiments, is one of the main goals of our analysis. In what follows, we introduce each type of dark matter interaction, setting the stage for the model-specific analysis in \cref{SEC:Dirac_DM,SEC:atomic-dm}. For the light dark photon masses of interest, the dark fermion $\chi=\{e',p'\}$ interactions with the electrons and protons in the detector are long-ranged, and the experimental bounds for light mediators apply. This is in contrast with elastic dark atom scattering, which proceeds through a short-range interaction as long as the dark-atom binding force is stronger than the dark fermion interactions with electrons or protons. In this latter case, bounds on dark matter scattering through contact interactions apply.

\paragraph{Dark fermion-nucleon scattering} 
The non-relativistic cross section for spin-independent dark matter-nucleus scattering can be written as \cite{Kaplinghat:2013yxa}
\begin{align}
    \sigma_{\chi N}(q^2) = \sigma_{\chi n} \vert_{q^2 = 0}\, A^2 \left(\frac{\mu_{\chi N}}{\mu_{\chi n}}\right)^2 \frac{m_d^4}{(m_d^2 + q^2)^2}\, F_N^2 (q^2)\,,
\label{ref_proton_scat}
\end{align}
Here $A$ is the mass number of the nucleus, $\mu_{\chi N}$ ($\mu_{\chi n}$)  the dark matter-nucleus (-nucleon) reduced mass, $F_N(q^2)$ the nuclear form factor and $q \equiv |\vec{q}|$ is the momentum exchange in non-relativistic scattering. The typical momentum exchange in dark matter-nucleus scattering is $q \approx 2 \mu_{\chi N} v$, with a local dark matter velocity of $v\approx 10^{-3}$. The  scattering cross section per nucleon $n$ at zero momentum exchange $q^2$ is
\begin{align}
    \bar \sigma_{\chi n} \equiv \sigma_{\chi n} \vert_{q^2 = 0} = \frac{16\pi \alpha \alpha_d \epsilon^2 \mu_{\chi p}^2}{m_d^4} \left(\frac{Z}{A}\right)^2,
    \label{ref_proton_scat0}
\end{align}
where $\alpha = e^2/4\pi$ is the electromagnetic fine-structure constant. The factor $(Z/A)^2$ takes into account that the dark-photon mediator couples only to the protons inside the nucleus, and we have averaged over all nucleons. For the most abundant xenon isotope in the XENONnT experiment, the charge and mass numbers are $Z = 54$ and $A = 131.29$ and $(Z/A) \approx 0.41$.

For a heavy mediator $m_d^2 \gg q^2$, the only momentum-dependence of the cross section is in the form factor $F(q)^2$, and experimental results are interpreted in terms of the reference cross section $\bar \sigma_{\chi n}$. For the light mediator, there is an additional factor $m_d^4/q^4$ in \cref{ref_proton_scat} from the propagator, and the $m_d$ dependence drops out of the cross section. To reflect this, the experimental bounds for light mediators are given in terms of the combination $\bar \sigma_{\chi n} m_d^4$, which is independent of the mediator mass.

\paragraph{Dark fermion-electron scattering} 
The dark matter-electron scattering cross section is given by
\be
\sigma_{\chi e}(q^2) = \overline{\sigma}_{\chi e}\, F_{e}^2(q^2),\qquad \text{with } F_{e}(q^2) = \frac{m_d^2 + q_{\rm ref}^2}{m_d^2 + q^2}.
\ee
The form factor $F_{e}(q^2)$ takes account of the momentum dependence of the scattering process;  $q_{\rm ref} = \alpha m_e$ is a typical reference momentum for experiments probing dark matter-electron scattering. 
For a heavy mediator, $F_e(q^2) = 1$, while for a light mediator, $F_e(q^2) = q_{\rm ref}^2/q^2$. In both cases experiments report their results in terms of the reference cross section $\bar\sigma_e$
given by~\cite{Knapen:2017xzo}
\begin{align}
    \overline{\sigma}_{\chi e} = 16\pi \alpha \alpha_d \epsilon^2\frac{\mu_{\chi e}^2}{(m_d^2 + q_{\rm ref}^2)^2}\,,
    \label{ref_elec_scat}
\end{align}
where $\mu_{\chi e}$ is the dark fermion-electron reduced mass.  

For Atomic DM there is the possibility of measuring both the dark electrons and protons, see \cref{SEC:atomic-dm}. 

\paragraph{Dark atom scattering}
Due to its net dark charge neutrality, the interaction between dark atoms and visible matter is effectively short-ranged.
The cross section for dark atom scattering is therefore much smaller than for unbound dark fermions. For small kinetic mixing, the interaction potential between the dark atom and a proton inside the detector material is much smaller than the interatomic potential, and can be treated as a perturbation. In this limit the cross section for elastic dark atom-proton scattering is
\begin{align}
    \sigma^{\text{el}}_{H' p} 
    =  4\pi\, \alpha_d\, \alpha\, \epsilon^2 \mu^2_{H' p}\, a_b^4\, \frac{(R -1)^2}{(R+1)^2}\,.
    \label{eq:sigma_Hf_el}
\end{align}
Here, $a_b = (\alpha_d \mu_{H'})^{-1}$ is the Bohr radius of the dark atom, and $\mu_{H' p}$  is the reduced mass of the dark atom and the proton.   In the limit $ R = m_{p'}/m_{e'} \gg 1$, \cref{eq:sigma_Hf_el} agrees with \cite{Cline:2012is}. In \cref{sec:app1} we explain our calculation of \cref{eq:sigma_Hf_el}, which extends the predictions for dark atom-nucleon scattering from \cite{Cline:2012is} beyond the regime of $R \gg 1$.

Elastic dark atom-proton scattering is a short-range process, in the sense that the force probes the charge distribution within the atom. The leading contribution to the cross section is momentum-independent and can be treated as a contact interaction. 
We define the dark atom-nucleon cross section as
\begin{align}
  \sigma_{H'n}^{\rm el} = \left(\frac{Z}{A}\right)^2 \sigma_{H'p}^{\rm el}\,,
\end{align}
which can be compared with experimental bounds on elastic dark matter scattering through a contact interaction by identifying $\sigma_{H'n}^{\rm el} = \bar\sigma_{n}$.

In general, dark atom-nucleon scattering is dominantly elastic. For equal-mass constituents, $ R \approx 1$, however, elastic scattering \cref{eq:sigma_Hf_el} vanishes, because the average charge distribution of the dark atom is zero. In this case, inelastic scattering due to spin-orbit coupling between the proton and the constituents of $H'$ becomes important. For $R=1$, the inelastic cross section involving hyperfine transitions of dark atoms is \cite{Cline:2021itd,Cline:2012is}
\begin{align}
\label{eq:sigma_Hf_inel}
    \sigma^{\text{inel}}_{H'p} \sim 16\, \alpha_d\,\alpha\,\epsilon^2 \left(\frac{m_p}{m_p+m_{H'}}\right)^2 \frac{v^2}{q^2}\,.
\end{align}
As $v^2/q^2\sim 1/m_{H'}^2$, we treat the inelastic cross section as momentum-independent.  For non-perturbative dark gauge couplings, the elastic cross section dominates and $\sigma^{\text{inel}}_{H'p} < \sigma^{\text{el}}_{H'p}$, except in the limit of degenerate constituent masses, $(R-1) \lesssim 0.1 \alpha_d^2$.

Dark atom-electron scattering can be treated analogously to dark atom-proton scattering. The cross sections for elastic and inelastic scattering are obtained by replacing $p \to e$ in \cref{eq:sigma_Hf_el} and \cref{eq:sigma_Hf_inel}.

\paragraph{Sensitivity of current experiments} We compare predictions of dark fermion and dark atom scattering with existing searches at direct detection experiments. For dark fermion scattering mediated by a light dark photon, we use the experimental bounds derived for massless or light mediators, which provide a good approximation in the relevant parameter range. For elastic dark atom scattering, we compare our predictions with searches for dark matter scattering through heavy mediators, that is, contact interactions.

For dark matter-nucleon scattering through a light mediator, the current bounds lie around $\bar\sigma_n \lesssim 4\cdot 10^{-37}\,\text{cm}^2$ at $m_{\rm DM} = 100\,$MeV and $m_{\rm DM} = 30\,$GeV, while the sensitivity is about an order of magnitude lower for DM masses around a few GeV. The most sensitive searches are by PandaX-4T~\cite{PandaX:2023xgl} for $m_{\rm DM} \in [0.03,2]\,$GeV with an S2-only analysis including the Migdal effect and taking the mass of light mediator $m_d =  0.1\,$MeV; by XENON1T~\cite{XENON:2019zpr} for $m_{\rm DM} \in [2,3]\,$GeV through the Migdal effect; by PandaX-4T~\cite{PandaX:2023xgl} for $m_{\rm DM} \in [3,10]\,$GeV as well as PandaX-II~\cite{PandaX-II:2018xpz} for $m_{\rm DM} \in [10,1000]\,$GeV, interpreting both S1+S2 analyses for $m_d = 1\,$MeV. The typical momentum exchange exceeds the mediator mass if $m_{\rm DM}/{\rm GeV} > 0.5  (m_d /{\rm MeV})$, for $m_{\rm DM}$ smaller than the xenon mass. This is satisfied in both of these searches, especially towards the upper range for the DM mass, and the bounds apply to lighter mediator masses as well. We will use these searches to constrain dark fermion-nucleon (identifying $\bar\sigma_{\chi n} = \bar\sigma_n$) scattering in both models in \cref{sec:direct_detection,SEC:direct-detection}.

Dark fermion-electron scattering through a light mediator is currently best probed by the SENSEI experiment~\cite{SENSEI:2023zdf}. SENSEI is sensitive to dark matter scattering in the range $m_{\chi} \in [1,1000]\,$MeV, with a maximum sensitivity of $\bar\sigma_{ e} \lesssim 2\cdot 10^{-36}\,\text{cm}^2$ around $m_{\chi} = 10\,$MeV. This search will be relevant for us to constrain dark fermion-electron scattering (identifying $\bar\sigma_{\chi e} = \bar\sigma_e$) in both models.

For spin-independent dark matter-nucleon scattering through contact interactions, the current bounds range from $\bar\sigma_n \lesssim 10^{-37}\,\text{cm}^2$ at $m_{\rm DM} = 100\,$MeV to $\bar\sigma_n \lesssim 4\cdot 10^{-47}\,\text{cm}^2$ at $m_{\rm DM} = 30\,$GeV. The most sensitive searches are by PandaX-4T~\cite{PandaX:2023xgl} for $m_{\rm DM} \in [0.03,1.3]\,$GeV with an S2-only analysis including the Migdal effect and using the heavy-mediator result for $m_d \ge 1\,$GeV; by Dark Side-50~\cite{DarkSide:2022dhx} for $m_{\rm DM} \in [1.3,3]\,$GeV through the Migdal effect; by PandaX-4T~\cite{PandaX:2023xgl} for $m_{\rm DM} \in [3,10]\,$GeV with an S1+S2 analysis; and by XENONnT~\cite{XENON:2023cxc} for $m_{\rm DM} \in [10,1000]\,$GeV. In~\cref{SEC:direct-detection} we will use these searches to constrain dark atom-nucleon scattering (identifying $\sigma^{\rm el}_{H' n} = \bar\sigma_n$) for Atomic Dark Matter.

Dark matter-electron scattering through contact interactions could in principle probe dark atom-electron scattering. However, current searches are not sensitive to Atomic Dark Matter within the astrophysically viable parameter space, see \cref{SEC:direct-detection,SEC:aDM_discussion}. 

\paragraph{Dark photon absorption}
Dark photon absorption by bound electrons in a detector is similar to photo-electric absorption. 
Stars emit a flux of dark photons with an energy set by the photon plasma mass, provided that the dark photon mass is smaller than the plasma frequency. For the Sun, this energy falls within the sensitivity range of direct detection experiments, and  dark photon absorption offers a competitive probe of kinetic mixing. 

The XENON1T collaboration has performed a search for an ionization (S2-only) signal~\cite{XENON:2019gfn}, which is sensitive to dark photon absorption. The analysis focuses on dark photons in the mass range $m_d = 0.186 - 1\,$keV. The bounds can be applied to smaller masses as well~\cite{An:2020bxd}, as they depend on the energy (rather than the mass) spectrum of the dark photon,
which is mostly determined by the plasma frequency.

We have extrapolated the results from Ref.~\cite{An:2020bxd} down to $m_d \sim 10^{-4}$ eV. Here we used that as the dark photon mass drops below the plasma mass, it decouples and the production rate in the Sun decreases as $m_d^2$. For the mass range $10^{-4}$ eV $\lesssim$ $m_d \lesssim 10\,$eV, the extrapolated results from XENON1T's ionization search translate into an upper bound on kinetic mixing
\be\label{eps_Sun}
\eps < 10^{-12}\,\frac{\mathrm{eV}}{m_d}\,.
\ee
For larger masses the numerical results of~\cite{XENON:2019gfn,An:2020bxd} can be used.
Lighter dark photons are excluded by bounds from spectral distortions of the CMB  \cite{McDermott:2019lch,Caputo:2020bdy,Arsenadze:2024ywr,Chluba:2024wui}, Cavendish-like experiments \cite{Kroff:2020zhp}, helioscopes~\cite{Betz:2013dza,Povey:2010hs,Parker:2013fxa,Romanenko:2023irv}, and  light-shining-through-the-wall (LSW) experiments \cite{Betz:2013dza,Romanenko:2023irv}.

If the dark gauge symmetry
is broken by a dark Higgs mechanism, strong -- though model-dependent -- bounds exclude the small $m_d$ mass region \cite{An:2020bxd,Cline:2024wja}. If kinetic mixing arises from integrating out a heavy degree of freedom, as was assumed in the derivation of \cref{higgs_bound},  stellar cooling limits the dark photon mass to   $m_d \gtrsim 2 \times 10^{-3}\,$eV. 

\paragraph{Sensitivity of future experiments} Current efforts to improve the sensitivity to dark matter scattering and particle absorption at direct detection experiments have two targets: smaller dark matter interaction rates; and smaller energy deposits, allowing to probe smaller dark matter masses.

Within the near future, existing experiments are expected to probe dark matter-nucleon scattering through contact interactions down to and into the so-called neutrino fog, i.e., background from neutrino scattering~\cite{Akerib:2022ort}. First indications for coherent neutrino-nucleus scattering have recently been reported by the XENON-nT and PandaX-4T collaborations~\cite{XENON:2024ijk,PandaX:2024muv}. The next generation of liquid-xenon-based experiments, including DARWIN/XLZD~\cite{Aalbers:2022dzr} and PandaX-30T~\cite{Wang:2023wrr}, will be sensitive to even smaller scattering rates. For Atomic Dark Matter, this means an improved sensitivity to the dark atom-nucleon scattering cross section by about an order of magnitude until reaching the neutrino fog. For Dirac Dark Matter, no such prediction is possible, since the background from the neutrino fog for dark matter-nucleon interactions through a light mediator has not been determined. We encourage such a light-mediator study, which would give valuable input for a broad class of models predicting dark matter scattering beyond contact interactions.

For sub-GeV dark matter, many new avenues exist to improve the sensitivity to dark matter-electron scattering and particle absorption~\cite{Essig:2022dfa}. Depending on the ability to reduce existing background and to lower the detection threshold, improved sensitivities up to several orders of magnitude in interaction strength and mass are predicted. For Dirac Dark Matter, we estimate that dark matter-electron scattering cross sections down to the neutrino fog can be probed in the next decade, see \cref{sec:discussion}. For Atomic Dark Matter, this improvement implies that smaller scattering rates for both dark electron and dark proton scattering can be probed. In both models, improving the sensitivity to small interaction rates is more important than to small masses, which are excluded by astrophysical bounds on dark matter self-interactions.

\subsection{Relic density}

We fix the dark matter abundance to the observed value of $\Omega_{\rm DM} h^2 = 0.120$~\cite{Planck:2018vyg}, but do not impose an explicit production mechanism as this generically introduces additional model dependence. Nevertheless, it is useful to discuss some possibilities, as this can give additional insights/bounds on the model parameters.

\paragraph{Dirac Dark Matter} The self-interaction constraints require the dark coupling $\alpha_d$ to be too small for efficient dark matter annihilation. If the dark matter sector was in thermal equilibrium with the SM sector at early times, the relic abundance from thermal freeze-out would be too large and overclose the universe. This assumes standard cosmology.

Dark matter can instead be produced via the freeze-in mechanism \cite{Hall:2009bx}, where SM particles annihilate via a virtual dark photon into dark matter. As the dark-sector interactions with the Standard Model are weak, the dark sector never reaches thermal equilibrium and dark matter annihilation can be neglected. Assuming that no other dark fields or forces are present beyond the dark electron and dark photon, the freeze-in abundance in a standard cosmology has been calculated in Ref.~\cite{Chu:2011be,Hambye:2018dpi}.

Direct detection experiments are on the brink of probing the freeze-in parameter space for this minimal set-up \cite{SENSEI:2023zdf}, and it has been argued that by re-interpreting direct detection analyses the
freeze-in scenario  can even already be tested for dark matter in the mass range $m_Z/2 \lesssim m_\chi \lesssim 100\,$GeV \cite{Hambye:2018dpi}. In our analysis in \cref{SEC:Dirac_DM}, we will indicate benchmarks for which freeze-in yields the correct relic abundance. It should be kept in mind that this benchmark applies for the minimal model of Dirac Dark Matter, while a richer dark sector or a non-standard cosmology could change these predictions. 

\paragraph{Atomic Dark Matter}
Thermal freeze-out of the dark electrons and dark protons yields different relic abundances of these particles for $R>1$, which would result in a large fraction of ionized dark atoms at late times. To ensure that most dark electrons and protons combine to form neutral dark atoms, we assume a dark baryon asymmetry $\eta_{p'}$ between the dark protons and anti-protons, similar (and possibly related) to the baryon asymmetry in the visible sector \cite{Kaplan:2009ag,Zurek:2013wia,Petraki:2013wwa}. The dark anti-protons are efficiently annihilated for sufficiently large dark gauge couplings $\alpha_d > 0.0035\,(m_{p'}/100\,$GeV) \cite{Boddy:2016bbu}, which is satisfied in the parameter space of interest. The dark and visible sector are in equilibrium at early times, and we fix the temperature $\xi =0.3$ consistent with \cref{eq:xi}.

\section{Dirac Dark Matter}
\label{SEC:Dirac_DM}

The parameter space for Dirac Dark Matter interacting via a dark photon that interacts with the Standard Model through kinetic mixing was analyzed in Ref.~\cite{Knapen:2017xzo}. As we will review, the constraints from self-interactions on the dark gauge coupling severely limit the possibilities for direct detection. On the flip side, if we do detect Dirac DM, the self-interactions are always large and may impact small scale structure. We focus on the detection possibility of both dark matter scattering and dark photon absorption.

The dark matter mass range in reach of current direct detection experiments is $m_{e'} = 1 \,{\rm MeV}- 1\,{\rm TeV}$; a
dark photon absorption
 is possible in the range $m_d = 10^{-4} - 10^2\,$eV \cref{eps_Sun}.
We define the reference values 
\be
\bar m_{e'} \equiv \frac{m_{e'}}{\rm GeV}\,, \quad
\bar m_d \equiv \frac{m_d}{10^{-4}\,{\rm eV}}\,, \quad
v_b \equiv \frac{v}{3 \times 10^{-3}}\,.
\label{dimless}
\ee
Velocities $v$ are quoted in units of the speed of light unless explicit units of km/s 
are used. Further $v_b$ is defined such that the typical dark matter velocity in the Bullet Cluster is $v_b=1$, while for small-scale structure $v_b =10^{-2}$.

\subsection{Constraints from astrophysics and cosmology} 

Dirac dark matter is strongly constrained by the Bullet Cluster and small-scale structure observations.
The transfer cross section for self-interactions in the  Born regime is \cite{Feng:2009hw,Tulin:2013teo} 
\be
\sigma^{\rm Born}_T =\frac{2\pi}{m_d^2} \beta^2\[ \ln(1+\R^2)-\frac{\R^2}{1+\R^2}\]
\approx \frac{2\pi}{m_d^2} \beta^2  \times \left\{ \begin{array}{ll} \frac12 \R^4, & \quad \R \ll 1 \\
\ln( \R^2)-1,& \quad \R\gg1.
\end{array} \right.
\ee
Here $\beta$ is the ratio of potential over kinetic energy at the interaction range of the effective Yukawa potential $r \sim m_d^{-1}$, and $\R$ is the interaction range over the dark matter de Broglie wavelength:
\begin{align}
\beta &= \frac{2\alpha_d m_d}{m_{e'} v^2} \approx 0.2 \times 10^{-7}\, \frac{\alpha_d \bar m_d}{\bar m_{e'} v_b^2 } &
\R &=\frac{m_{e'} v}{m_d} \approx 3 \times 10^{10}\, \frac{\bar m_{e'} v_b}{\bar m_d} .
\label{betaR}
\end{align}
The perturbative Born approximation breaks down for $\frac12 \R^2 \beta = (m_{e'}\alpha)/m_d \gg 1$. The photon mass can be neglected for $\R \gg 1$, and the interaction is well described by the Coulomb force. Then the non-perturbative cross section is well described by the classical approximation. For $\beta \lesssim 0.1$ this gives \cite{Tulin:2013teo}
\begin{equation}
   \sigma_T^{\mathrm{clas}} 
       \frac{4\pi}{m_d^2}\, \beta^2 \ln( \beta^{-1}),  \qquad \beta \lesssim 10^{-1},
\end{equation}
both  for attractive particle-antiparticle and repulsive (anti)particle-(anti)particle interactions. In our analysis, we will not disentangle their respective effects on small-scale structure observables, and our bounds on $\alpha_d$ will be subject to $\O(1)$ uncertainties. The Born and classical approximations for the cross section differ by $\O(1)$ in the region where both are valid. 
For simplicity, we will use the Born cross section for
the analytical estimates in this section; in our numerics we switch to the classical approximation for $\frac12\R^2\beta >1$.

For $\beta \ll 1$ and $\R \gtrsim 1$, constraints on dark matter self-interactions from the Bullet Cluster~\eqref{bullet} set an upper bound on the dark gauge coupling, 
\be
{\rm Bullet\ Cluster}: \qquad \alpha_d \lesssim 
2 \times 10^{-5}\, \bar m_{e'}^{3/2} v_b^{2}\( \frac{25}{\ln \R^2}\)^{1/2} \qquad  {\rm for} \;\; \R \gtrsim 1.
\label{bullet_Dirac}
\ee
Due to the strong velocity dependence of the transverse cross section, $ \sigma_T/m_{e'} \propto v^{-4}$, small-scale structure bounds \cref{small_scale} with $v_b =  10^{-2}$ 
give a constraint on $\alpha_d$ that is a factor $v_b^2 \sqrt{c^{\rm ss}} =10^{-4} \sqrt{c^{\rm ss}}$ stronger than from the Bullet Cluster. For the small gauge couplings required by self-interactions,
the approximation $\beta \ll 1$ is indeed valid for the dark sector masses of interest, see \cref{betaR}.

For small dark photon masses, the dark electron effectively behaves as millicharged dark matter in stellar environments, and stellar cooling bounds give the constraint \cref{stellar_cooling}
\be
\alpha_d\, \eps^2\lesssim 10^{-18}, \qquad m_{e'} = 0.01-1\,{\rm MeV}.
\label{milicharged}
\ee
For lighter masses the bounds are even stronger. For large kinetic mixing close to experimental bounds, the self-interaction constraint \cref{bullet_Dirac} gives the most stringent bound on the dark gauge coupling. However, if Dirac Dark Matter is only a subdominant fraction of the total dark matter, small-scale structure bounds disappear and stellar cooling becomes relevant.

Dark matter much heavier than the dark-sector temperature at the time of recombination behaves as cold dark matter and bounds from dark acoustic oscillations do not apply. Dark electrons decouple from the dark thermal bath a little before the interaction rate becomes smaller than the expansion rate of the universe. This happens for dark-sector temperatures $T_d \approx 10^{-6} \alpha_d^{-1} \bar m_{e'}^{3/2}\,{\rm eV}$ \cite{Bringmann:2016ilk}.
Given the strong bound on $\alpha_d^{-1} \bar m_{e'}^{3/2}$ dictated by the Bullet Cluster~\cref{bullet_Dirac},  $T_d$ lies well above the recombination temperature $T_{\rm rec}\approx {\rm eV} $ for dark matter masses within the range of direct detection experiments. As a result, the DAO constraints do not significantly impact our parameter space.

\subsection{Direct detection} 

Dirac dark matter can scatter off both electrons and nucleons in the detector.
The effective cross sections for scattering with nucleons and electrons that are reported by experimental searches are given in \cref{ref_proton_scat0,ref_elec_scat}, respectively. 
We parameterize the  dark matter self-interaction bound \cref{bullet_Dirac} (for $\R \gg 1$)  as 
\be
\alpha_d =  2 \times 10^{-5} \, \bar m_{e'}^{3/2} x\,,
\label{xy}
\ee
with $x \lesssim 1 \;(10^{-4}\sqrt{c^{\rm ss}})$ to satisfy the constraints from the Bullet Cluster (small-scale structure).
The DM-nucleon and DM-electron reference cross sections can then be written as
\begin{align}
\bar\sigma_{e'n} m_d^4 &= 16 \pi\, \alpha \,\alpha_d\, \mu_{e'p}^2\, \eps^2  \left(\frac{Z}{A}\right)^2 
 =
 6 \times 10^{-38}\,{\rm cm}^2 \,\text{MeV}^4\, x\, \bar m_{e'}^{3/2}\( \frac{\eps}{10^{-8}}\)^2\(\frac{\mu_{e'p}}{m_p}\)^2  
 \nn \\
{\bar \sigma_{e'e}} &\approx \frac{16 \pi \alpha \alpha_d  \eps^2 \mu_{e'e}^2 }{(\alpha m_e)^4}  = 5 \times 10^{-34} {\rm cm}^2\, x\, \bar m_{e'}^{3/2} \( \frac{\eps}{10^{-8}}\)^2 \(\frac{\mu_{e'e}}{m_e}\)^2,
\label{eq:sigma-e}
\end{align}
where $m_p \approx 1\,{\rm GeV}$ and $m_e \approx 0.5\,{\rm MeV}$ are the proton and electron mass.
Here $\eps \lesssim 10^{-8}$ is the maximum experimentally allowed mixing for dark photon masses $m_d \lesssim 10^{-4}\,$eV, see \cref{eps_Sun}, with saturation corresponding to the current absorption bounds from Xenon1T, see \cref{sec:direct_detection}.
The numerical factor in each expression gives the maximum cross section if constraints from absorption, the Bullet Cluster, and lab searches are satisfied. 

The relative rates for dark photon absorption over dark matter scattering scale as $\epsilon^2/\alpha_d$. For near-maximum kinetic mixing and a perturbative dark coupling strength, dark photon absorption from the Sun can occur simultaneously with dark matter scattering. For dark matter masses $m_{e'} \gtrsim 6\,$GeV, absorption and scattering can be experimentally distinguished, because dark matter scattering is currently more sensitively detected using a combined S1-S2 signal. For smaller dark matter masses, simultaneous absorption and scattering off electrons or nucleons can only be distinguished through the energy spectrum in case a signal is observed.

\subsection{Discussion}\label{sec:discussion}

\begin{figure}[t!]
   \centering
   \includegraphics[height=0.48\textwidth]{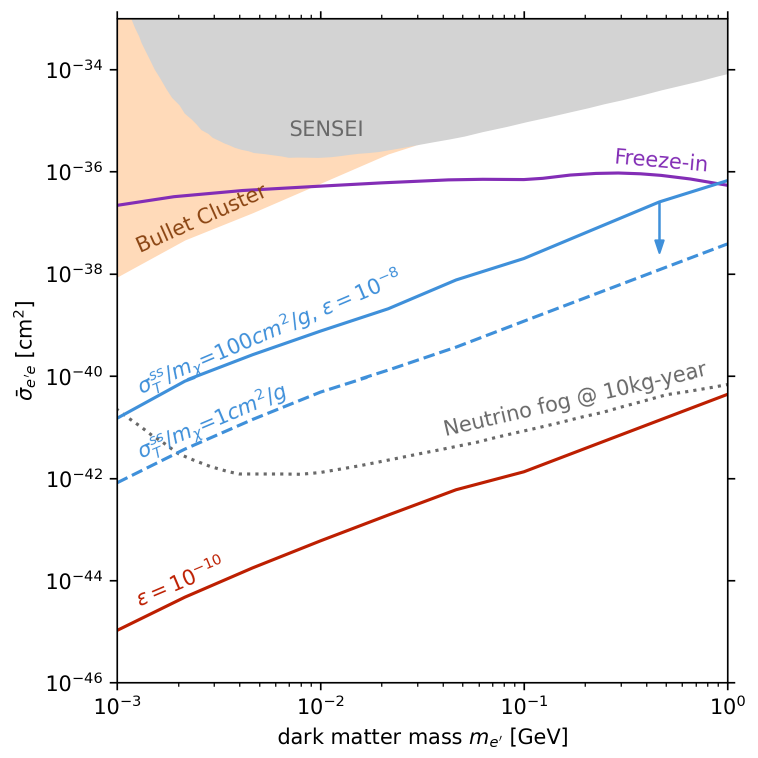}\hspace*{0.4cm}
   \includegraphics[height=0.48\textwidth]{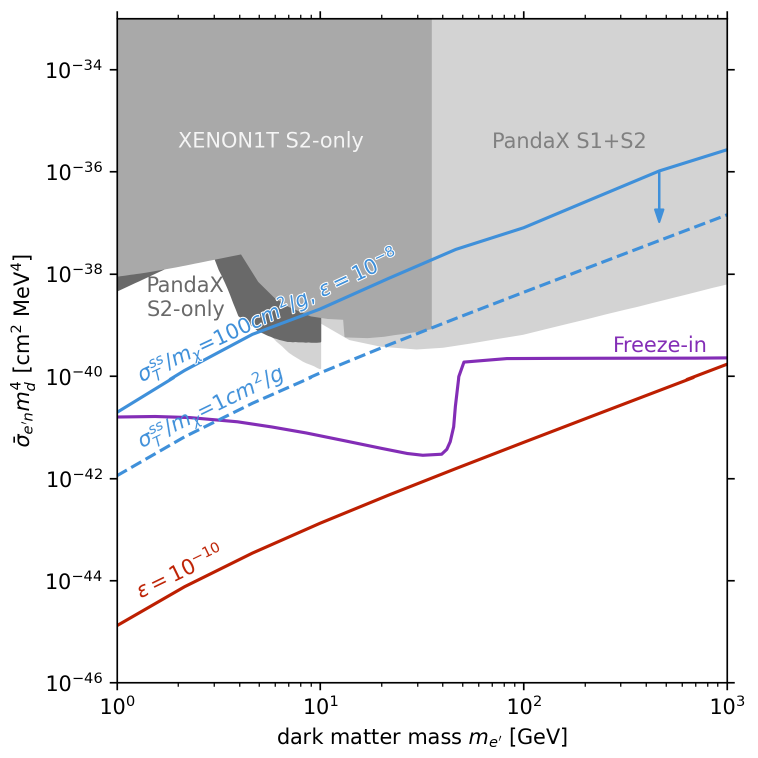}
   \caption{Dirac Dark Matter scattering cross sections with electrons $\bar \sigma_e$ (left) and nucleons $\bar{\sigma}_n$ (right). Direct detection limits from SENSEI \cite{SENSEI:2023zdf}, XENON1T~\cite{XENON:2019gfn,XENON:2019zpr}, and PandaX~\cite{PandaX-II:2018xpz,PandaX:2023xgl} are shown as grey-shaded regions. The region excluded by the Bullet Cluster~\cite{Randall:2008ppe,Markevitch:2003at} for $\epsilon = 10^{-8}$ is shaded orange. The maximum cross sections allowed by small-scale structure constraints, $\sigma_T^{ss}/m_{e'} \equiv \sigma_T(v=10 {\rm km/s})/m_{e'} \leq 100\,$cm$^2/$g (solid lines) and $\leq 1\, $cm$^2/$g (dashed lines), are shown for $\eps =10^{-8}$ (blue) and $\eps =10^{-10}$ (red). For reference, we show the neutrino fog at 10kg-year~\cite{Essig:2018tss}, where solar neutrino scattering starts to become a considerable background (dotted grey line). Parameter points that give the observed relic abundance through freeze-in \cite{Chu:2011be} are shown as purple lines.}
   \label{fig:result-Dirac}
\end{figure}

In \cref{fig:result-Dirac}, we show the available parameter space consistent with all constraints discussed in the previous subsections. The left panel shows the DM-electron reference cross section $\bar \sigma_{e'e}$  for dark matter masses in the MeV-GeV range. The right panel shows DM-nucleon scattering in terms of $\bar \sigma_{e'n}$, which is relevant for GeV-TeV dark matter masses. In these mass ranges, constraints from millicharged particle searches \cref{milicharged}, the CMB or BBN are absent.

The scattering rates from \cref{eq:sigma-e} are proportional to $\alpha_d\, \eps^2$, and are maximized by sa\-turating the observational constraints on the dark gauge coupling and kinetic mixing. XENON1T's search for dark-photon absorption from the Sun sets a dark-photon-mass-dependent bound on kinetic mixing \cref{eps_Sun}.  The dark gauge coupling is severely constrained by the self-interaction bound \cref{xy}. In the left panel, the parameter region excluded by the Bullet Cluster \cite{Randall:2008ppe,Markevitch:2003at} is shaded orange; in the right panel, this region is hidden behind the direct detection limits.

If one further imposes the small-scale structure constraints, only the regions below the blue lines remain available (indicated by arrows pointing downward). We show both $\sigma_T/m_{e'} = 1\,$cm$^2/$g (dashed lines), where self-interactions may address the small-scale structure problems of collisionless dark matter, and $\sigma_T/m_{e'} = 100$ cm$^2/$g (solid lines), which can be viewed as an upper bound. These cross sections correspond to $x =10^{-4}$ and $x = 10^{-2}$ in \cref{xy}, respectively. 
For comparison, we also show the upper bound on the scattering rate that is consistent with $\sigma_T/m_{e'} = 100$ cm$^2/$g for $\eps =10^{-10}$.

Freeze-in scenarios produce the observed relic abundance along the purple lines, assuming a standard cosmology and dark electron production via a dark-photon mediator \cite{Chu:2011be,Hambye:2018dpi}. In the right panel, the jump around $m_{e'} = M_Z/2$ is due to the $Z$-boson resonance.

In the left panel of \cref{fig:result-Dirac}, we show the currently most stringent bounds on DM-electron scattering from SENSEI \cite{SENSEI:2023zdf}. Taking only the Bullet Cluster constraints into account, the maximum DM-electron scattering cross section~\cref{eq:sigma-e} is within the current sensitivity range of direct detection experiments,  and at the same time dark-photon absorption is possible. However, if the much stronger small-scale structure bounds hold up, the detector sensitivity must be improved by more than a factor of 100 to be able to detect dark electron scattering. According to the forecasts in \cite{Essig:2022dfa}, such an improvement is within reach of near-future experiments. In the presence of small-scale structure bounds, viable scattering rates are too small to obtain the observed relic density via freeze-in. 

Dark-photon absorption is independent of $\alpha_d$ and therefore largely unaffected by self-interaction constraints. With an improved sensitivity, direct detection experiments can probe kinetic mixing below $\epsilon = 10^{-8}$ through dark-photon absorption, keeping in mind that the absorption rate scales as $\epsilon^4$.
If we drop the requirement that dark-photon absorption is within experimental reach, we can consider a much smaller dark photon mass and push the mixing parameter up to  the largest mixing allowed by LSW experiments and CMB observations, $\eps \approx 10^{-6}$. Freeze-in becomes viable again, and DM-electron scattering (the rate scales $\propto \eps^2$) is observable in the near future, while small-scale structure constraints are satisfied. 

In the right panel of \cref{fig:result-Dirac}, we show the currently strongest bounds on DM-nucleon scattering from XENON1T~\cite{XENON:2019gfn,XENON:2019zpr} and PandaX-II~\cite{PandaX:2023xgl}. When maximizing $\alpha_d \eps^2$, both DM-nucleon scattering and photon absorption rates are within reach at current experiments for dark matter masses above about $3\,$GeV. Interestingly, in the mass range $1-6\,$GeV the sensitivity to scattering is dominated by S2-only searches \cite{XENON:2019gfn,XENON:2019zpr,PandaX:2023xgl}. If dark-matter scattering and dark-photon absorption both occur at detectable rates in this mass range, spectral information is needed to disentangle their signatures. Freeze-in sets an interesting target; for dark-matter masses just above the $Z$ pole, the relevant parameter space can be probed in the near future.

As direct detection experiments improve their sensitivity, the `neutrino fog' where the dark matter signal becomes statistically indistinguishable from coherent solar neutrino scattering coherently off nuclei will be crucial background. The neutrino fog depends on the detector material and the exposure time. In the left panel of \cref{fig:result-Dirac}, we show the fog predictions for xenon and a 10kg-year exposure from~\cite{Essig:2018tss}. Additional dependence on the ionization efficiency would broaden this line. In recent $S_2$-only searches for the absorption of light bosons, solar neutrino scattering is already the dominant background. These analyses already see the onset of the fog~\cite{Essig:2018tss}. To  our knowledge, no equivalent determination of the neutrino fog for DM-nucleon scattering mediated by a light mediator exists. 

Throughout the discussion, we have assumed that the dark electron makes up all of the dark matter. 
If it is only a subdominant component of the total relic abundance, a fraction of less than $1-10\%$, self-interaction constraints are absent \cite{Boddy:2016bbu,Knapen:2017xzo}. The dark matter scattering rates are suppressed by the same fraction, 
but this can be compensated by larger gauge couplings, which are only constrained to non-perturbative values $\alpha_d \lesssim 1$.
For large dark couplings and depending on the dark matter mass, the cross section for both DM-electron and DM-nucleon scattering can be within reach at current experiments.  We note, however, that this scenario only works if the relic density of dark protons is negligible. Otherwise, due to the large dark gauge coupling, the dark electrons and protons bind into dark atoms, which changes the phenomenology.

\section{Atomic Dark Matter}
\label{SEC:atomic-dm}

Atomic Dark Matter has a richer direct detection phenomenology than Dirac Dark Matter, due to the presence of both dark atoms and dark constituents in the galactic halo. We first review the phenomenology of Atomic Dark Matter in cosmology and astrophysics and determine the viable parameter space. We then discuss the prospects for direct detection of all four particle species in this model: dark atoms, dark electrons and dark protons from ionized dark atoms, and dark photons. The interplay of dark-atom and dark-constituent scattering with electrons and protons inside the detector material crucially depends on model parameters such as the mass ratio between the constituents. As for Dirac Dark Matter, we will answer the question whether dark matter scattering and dark photon absorption can simultaneously be observed at current and future experiments.

\subsection{Constraints from astrophysics and cosmology}
\label{sec:aDM_cosmo}

\paragraph{Self-interactions} In the Atomic Dark Matter model, self-interactions are present between the constituents (dark protons and dark electrons), among the dark atoms, and between dark atoms and constituents. As the dark atom is neutral under the dark $U(1)_d$ at large distances, self-interactions between dark atoms are much weaker than between the constituents.
If the fraction of ionized dark atoms during structure formation is small, the impact of constituent scattering is negligible and the bounds from small-scale structure are significantly relaxed compared to the case of Dirac Dark Matter. 

At dark recombination, not all constituents combine to form neutral dark atoms via the process $e' + p' \to H' + \gamma_d$. The fraction of ionized atoms remaining after the freeze-out of this process, estimated in \cite{Cline:2013pca,Kaplan:2009de}, should satisfy\footnote{We differ in notation with \cite{Cline:2021itd}, which does not include the dark electron number density in the denominator. We follow the parametrization from \cite{Kaplan:2009de}, and for $f_i\ll1$ both parametrizations are the same.}
\begin{align}
    f_i \equiv n_{e'}/(n_{e'} + n_{H'}) \approx \min \left[1, 10^{-10}\, \xi\, \alpha_d^{-4} R^{-1} \left(\frac{m_{H'}}{\GeV}\right)^2\right]\leq 0.01-0.1
    \label{ion}
\end{align}
to avoid the strong bounds on self-interactions of the constituents. We  use $f_i\leq 0.01$ in our numerical scans.

Additionally, dark matter in halos can heat during structure formation as overdensities collapse and virialize. If the virial temperature $T_{\rm vir}$ exceeds the binding energy $B_{H'}$, the dark atoms reionize, which increases the ionization fraction $f_i$ at late times \cite{Ghalsasi:2017jna}. Requiring the absence of reionized dark atoms through $T_{\rm vir} \lesssim 0.1 B_{H'}$, thus avoiding the strong self-interaction constraints for constituents, sets a lower bound on the dark coupling \cite{Ghalsasi:2017jna,Cline:2021itd}

\begin{align}
    \alpha_d \gtrsim 1.4 \times 10^{-3} \sqrt{R}\qquad \text{for }R \gg 1 \label{reion}.
\end{align}
Cooling in the dark halos through Bremsstrahlung, Compton scattering and atomic cooling is effective in parts of the parameter space~\cite{Fan_2013,Ghalsasi:2017jna}. The extra cooling can lower the temperature and thus the fraction of reionized atoms, so that small-scale structure bounds could be satisfied even for smaller dark couplings.  A careful treatment of these processes requires  numerical simulations, and we stick to dark couplings that satisfy~\cref{reion} for simplicity.

If both \cref{ion,reion} are satisfied, the ionization fraction is small and only dark hydrogen
self-interactions can affect structure formation significantly. The cross section for non-relativistic dark hydrogen scattering can be calculated numerically in a partial-wave expansion in the Born-Oppenheimer approximation, which requires $R\gg 1$ \cite{Cline:2013pca}. The result depends on three quantities, which can be taken as the Bohr energy $\eps_b =\alpha^2 \mu_{H'}$, the Bohr radius $a_b = (\alpha_d \mu_{H'})^{-1}$, and the mass ratio $R = m_{p'}/m_{e'}$, defined in \cref{R,bohr}. When expressed in Bohr units, only the $R$-dependence remains.

In the non-perturbative regime, the $H' - H'$ scattering cross section exhibits a series of resonance peaks as a function of  $R$. Away from the resonances the cross section in Bohr units, as indicated by the superscript, can be fitted by \cite{Cline:2013pca}
\begin{equation}
    \sigma_T^{\rm Bohr} (E) = \frac1{a_0 (R) + a_1(R) E + a_2 (R) E^2} \label{eq:xsect-HH}
\end{equation}
for fitting coefficients $a_i(R)$ and center-of-mass scattering energies $E = f(R) ({v}/{\alpha_d})^2 $ (in Bohr units), with $f(R)$ defined in \cref{R}. We interpolate the coefficients $a_i(R)$ for arbitrary $R$ based on the discrete fit in \cite{Cline:2013pca}. For $R \lesssim 100$, the coefficients are approximated by $\{a_0,a_1,a_2\} \approx \{6\times 10^{-3},\,0.3,\,0.05 \}$ with $\O(1)$ uncertainties; for larger $R$ the fit becomes increasingly less accurate. For this reason, we restrict our parameter space to $R\leq 4000$.

For $a_0 > a_1 E$, the s-wave contribution dominates in the denominator of \cref{eq:xsect-HH}. The self-interaction cross section is thus approximately velocity-independent for
\be
 f(R) \approx R \lesssim R_{\rm crit} \equiv 2.4 \times 10^3\alpha_d^2 \( \frac{10^3\,{\rm km/s}}{v}\)^2.
\label{Rcrit}
\ee
For small mass ratios $R$ and large couplings $\alpha_d$, the binding energy exceeds the kinetic energy and the cross section becomes geometrical, $\sigma_T \propto a_b^2$. In this regime, increasing the coupling tightens the constituents inside the dark atoms and thus reduces the interaction cross section.  For large dark couplings $\alpha_d \gtrsim 10^{-2}$,  small-scale structure bounds ($v=10\,$km/s) are determined by the s-wave contribution $a_0$ for $R \lesssim 10^4$, whereas the Bullet Cluster bound ($v=10^3\,$km/s) is sensitive to momentum-dependent corrections for $R>R_{\rm crit}$.

In our numerical scans of the parameter space, we use the full cross section \cref{eq:xsect-HH} to determine the bounds from the Bullet Cluster and small-scale structure. An analytic approximation can be obtained by restricting the cross section to the s-wave contribution $\sigma_T \approx \sigma_T(E = 0)$ for $R<R_{\rm crit}$.
Away from the resonances, one finds\footnote{The s-wave resonances are located at $R_n \approx -3.45 + 9.49 n + 7.74 n^2$ \cite{Cline:2013pca}.}
\begin{equation}
   \frac{\sigma_T(E = 0)}{m_{H'}} \approx 2.9\, \frac{\text{cm}^2}{\text{g}} \times f(R)^2 \(\frac{0.1}{ \alpha_d} \)^2 \(\frac{\text{GeV}}{m_{H'}}\)^{3}. 
   \label{self_swave}
\end{equation}
For the small-scale structure bound \cref{small_scale}, we use the very good approximation $\sigma_T(E=0)/m_{H'} \lesssim c^{\rm ss}{\text{cm}^2}/{\text{g}}$ with $c^{\rm ss}=1-10^2$ in the parameter space of interest. 
For the Bullet Cluster bound \cref{bullet}, the p-wave contribution becomes important for 
 $R>R_{\rm crit}$ and the transfer cross section is approximately a factor $R_{\rm crit}/R$ weaker. Hence, the Bullet Cluster constraint can be approximated by 
$\sigma_T(E = 0)/m_{H'} \times {\rm min}(1, R_{\rm crit}/R)\lesssim {\text{cm}^2}/{\text{g}}$. For small $R$, where the cross section is velocity-independent, the Bullet Cluster gives the strongest constraint and it is hard to address small-scale structure problems with self-interactions.

\paragraph{Dark molecule formation}
In dark atom interactions, dark molecules $H_2'$ consisting of two dark atoms could form. Direct formation via $H'+H'\to H'_2 +\gamma_d$ proceeds through an electric quadrupole and is  suppressed~\cite{Cline:2013pca}. The most efficient way to produce $H'_2$ instead is to start with dark hydrogen-constituent scattering to first form $H_2'^+$ and $H'^-$, followed by subsequent scattering with dark hydrogen. The first step in this chain is curtailed by the small ionization fraction. The ratio of dark molecules $H'_2$ over dark atoms $H'$ is typically less than $10^{-6}$, unless dark recombination happens at redshifts much larger than  recombination in the Standard Model~\cite{Gurian:2021qhk}. For the range of dark-sector temperatures $\xi$ (see \cref{eq:xi}) and couplings $\alpha_d$ which are relevant in this work, this is not the case. We conclude that dark molecule formation is not effective for viable scenarios of Atomic Dark Matter.

\paragraph{Cosmic microwave background} 
Constraints from dark acoustic oscillations appearing imprinted on the CMB can be rephrased in terms of $\Sigma_{\rm{DAO}}$, see~\cref{Sigma_gen}. For Atomic Dark Matter \cite{Cyr-Racine:2012tfp,Cyr-Racine:2013fsa}
\begin{equation}
    \Sigma_{\rm{DAO}} \equiv 
        \frac{2\times10^{-9}}{\alpha_d} \frac{(1+R)^2}{R} \Big(\frac{m_{H'}}{\mathrm{GeV}} \Big)^{-7/6}, 
\end{equation}
which depends on the dark radiation - dark matter coupling strength $\alpha_d$, as well as the dark matter mass (heavier dark matter decouples earlier in the universe's history).  The CMB bound on $\Sigma_{\mathrm{DAO}}$ in \cref{Sigma_bound} is satisfied for all of the parameter space that is in agreement with self-interaction constraints.

\subsection{Direct detection}
\label{SEC:direct-detection}
Atomic Dark Matter can be partially ionized, as recombination does not capture all charged particles. The remaining constituents of ionized dark atoms scarcely populate the dark matter halo and can potentially be detected.\footnote{Millicharged dark constituents can be ejected from the galaxy by supernova shock waves \cite{Cline:2012is,Chuzhoy:2008zy,McDermott:2010pa}. In our set-up, millicharges only arise  if the SM photon obtains an effective plasma mass~\cite{Feldman:2007wj}. We expect that this effect is small in supernova winds, so that the ionized constituents remain in the galaxy until today.}

Direct detection experiments assume a single species of dark matter when reporting exclusion limits, that is, they assume an abundance $\rho_{\rm DM}/m_\chi$ rather than the correct value $f_i \,\rho_{\rm DM}/m_{H'}$.
To take this into account, we rescale our predictions for constituent-nucleon and constituent-electron scattering to the effective cross sections
\begin{align}
\label{eq:sig_eff}
 \bar \sigma_{\chi n,\rm eff} = \frac{m_\chi}{m_{H'}} f_i\, \bar \sigma_{\chi n},\qquad \bar \sigma_{\chi e,\rm eff} = \frac{m_\chi}{m_{H'}} f_i\, \bar \sigma_{\chi e}\,.
\end{align}
We have chosen to absorb the flux factor into the effective cross section, so that we can compare dark electron and dark proton scattering with the same experimental bounds on $\bar\sigma_{n}$ or $\bar\sigma_{e}$. The effective cross section for dark electron scattering is thus suppressed by a factor $m_\chi/m_{H'}\approx 1/(1+R)$ compared to dark proton scattering.
Given the small ionization fraction $f_i \ll 1$ required by structure formation, we neglect the cross section correction on dark atom scattering and set $1 - f_i \approx 1$. For a given set of model parameters, the ionization fraction $f_i$ is fixed through~\cref{ion}.

While the charged dark fermions couple more strongly to SM particles than the neutral dark atoms, their interaction rate is suppressed by the smaller abundance. It is therefore a priori not clear which particle species dominates the scattering signal. As we will see, different processes can dominate the signal in different parts of the parameter space.

\begin{figure}[t!]
   \centering
   \includegraphics[height=0.35\textwidth]{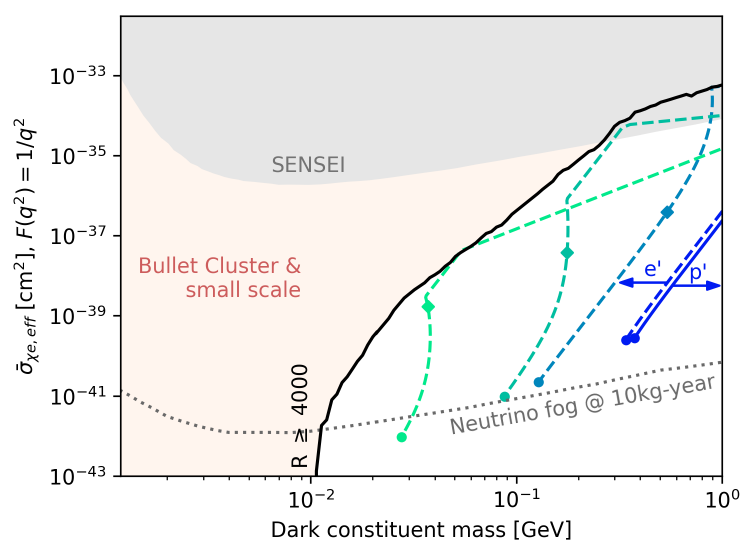}\hfill
   \includegraphics[height=0.35\textwidth]{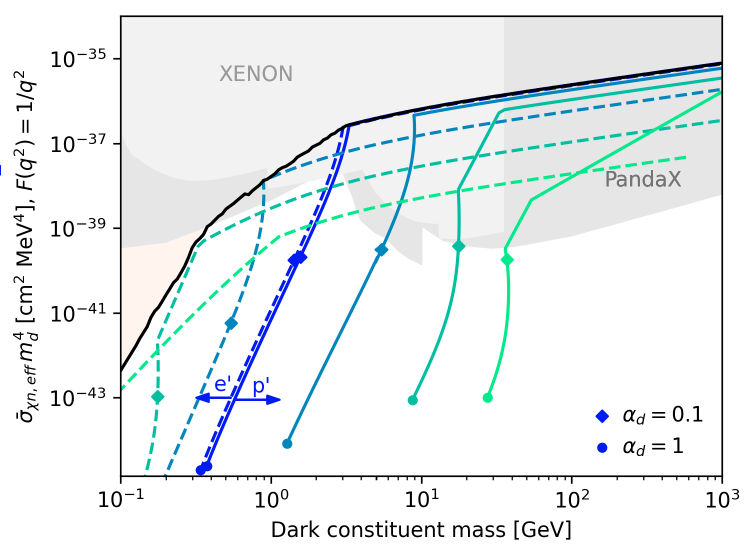}\\\vspace*{0.3cm}
   \includegraphics[height=0.35\textwidth]{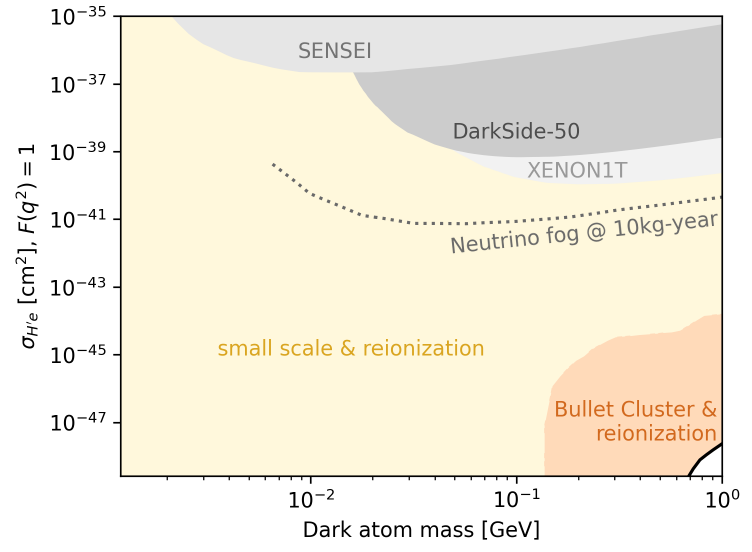}\hfill
   \includegraphics[height=0.35\textwidth]{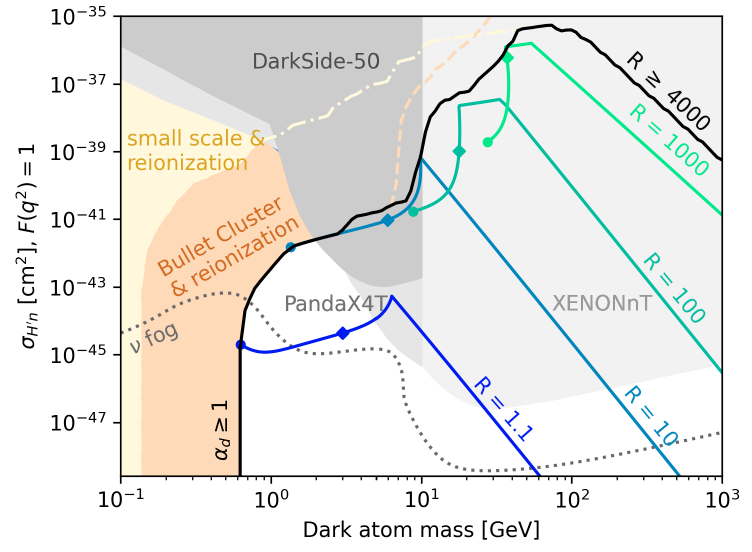}
   \caption{Sensitivity of current direct detection experiments to DM-electron scattering (left column) and DM-nucleon scattering (right column) of Atomic Dark Matter. Experimental results are shown for a massless mediator, relevant for dark electron or dark proton scattering through a light dark photon (top row), and for contact interactions, relevant for dark-atom scattering (bottom row). Current limits from SENSEI \cite{SENSEI:2023zdf}, DarkSide \cite{DarkSide:2022dhx,DarkSide:2022knj}, XENON \cite{XENON:2023cxc,XENON:2019zpr,XENON:2019gfn}, and PandaX \cite{PandaX:2023xgl,PandaX-II:2018xpz} are shown in grey; the corresponding neutrino fog \cite{Essig:2018tss,OHare:2021utq} as grey dotted lines. Scenarios that satisfy all constraints from astrophysics lie below the solid black lines. Predictions of dark atom and dark constituent scattering (blue-to-green colored lines) are shown for fixed mass ratios $R = \{1.1,10,100,1000\}$, kinetic mixing $\epsilon = 10^{-8}$, and $\alpha_d$ maximizing the cross section while satisfying astrophysical bounds. On these lines, dots (diamonds) of the same color correspond to $\alpha_d = 1$ ($\alpha_d = 0.1$) from self-interaction constraints and the same dark-sector parameters $m_{H'}$ and $R$. In the upper panels, dashed (solid) lines of constant $R$ at smaller (higher) masses correspond to dark electron (dark proton) scattering.
        \label{fig:dd-atomic-dm}}
\end{figure}

In \cref{fig:dd-atomic-dm}, we present the sensitivity of current direct detection experiments to Atomic Dark Matter. We indicate bounds on the parameter space due to a combination of non-reionization with small-scale structure (yellow), of non-reionization with the Bullet Cluster (orange), and of the Bullet Cluster with small-scale structure (light red). Predictions of dark atom and dark constituent scattering (blue-to-green colored lines) are shown for fixed kinetic mixing $\epsilon=10^{-8}$ and various mass ratios $R$. The choice of kinetic mixing maximizes the scattering cross section and restricts the dark photon mass below the meV scale to evade the bound on dark photon absorption by XENON1T, \cref{eps_Sun}. Since the momentum transfer in dark matter scattering is well above this scale, the dark photon mass plays no role in the displayed direct detection phenomenology.

Along the curves of constant $R$, the dark gauge coupling $\alpha_d$ varies in a non-trivial way, due to the interplay of several bounds from astrophysics. The curve ends at small masses as the non-perturbativity bound is reached $\alpha_d \leq 1$, indicated by the dot. As the mass is increased, the dark gauge coupling decreases and the cross section increases (hence an increasing iso-$R$-line); this behavior is dictated by self-interaction constraints. At larger masses the bound on the ionization fraction $f_i \leq 0.01$ dominates and $\alpha_d$ starts to increase. For large enough $R$, there is a small intermediate mass range in which the reionization constraint becomes important, and which demands a larger $\alpha_d$ than allowed by self-interaction constraints. This results in shallower slopes for dark constituent effective cross sections and negative slopes for dark atom cross sections. Due to the different $\alpha_d$ scaling of the most stringent constraint, the different mass regimes are separated by sudden breaks in the iso-$R$ lines.

Dark constituent interactions are mediated by a light dark photon and are long-ranged, whereas dark atom interactions can be treated as short-ranged. Since direct detection experiments are sensitive to this difference, the bounds on the effective cross section for constituent scattering and dark atom scattering cannot be compared one-to-one. Moreover, since the optimal dark coupling can be different between for dark atom and constituent scattering, it might be that predictions for constituent scattering that are within reach of an experiment are already excluded by searches for dark atom scattering by another search - and vice versa. To answer the question whether simultaneous scattering of dark atoms and constituents or simultaneous dark matter scattering and dark photon absorption occurs in current experiments, we refer the reader to \cref{SEC:aDM_discussion}.

To better understand the intricate relations between dark atom and constituent scattering, we analyze the parameter dependence of the various scattering processes in detail.

\paragraph{Dark fermion-electron scattering}
The effective scattering cross section for dark constituents off electrons can be expressed as (see \cref{ref_elec_scat})
\begin{align}
\label{eq:sig_eeff}
    \bar \sigma_{\chi e,\mathrm{eff}} 
   & = 2\times 10^{-30} \mathrm{cm}^2  \left(\frac{\epsilon}{10^{-8}}\right)^2  \left(\frac{\mu_{\chi e}}{m_e}\right)^2 \left(\frac{\alpha_d}{0.1}\right)
   \(\frac{m_\chi}{m_{H'}}\)f_i(m_H',R,\alpha_d). 
\end{align}
The only difference between dark electron and dark proton scattering is in the reduced mass $\mu_{\chi e}$, which is negligible for $m_{p'},m_{e'}\gg m_e$. The ionization fraction is also the same for dark electrons and dark protons.  However, experimental bounds on $\bar \sigma_{e}$ are extracted assuming that the scatterer makes up all of the relic dark matter abundance, and overestimate the $\chi$-abundance. This is corrected by an extra factor $f_i \,m_\chi/m_{H'}$ in the effective cross section, see~\cref{eq:sig_eff},  which favors the heavier dark protons over dark electrons. Moreover, for a given constituent mass $m_\chi$ in the sensitivity range of an experiment, that is, sub-GeV masses for electron scattering, the dark atom mass is a factor of $(1+R)$ larger for dark electrons than for dark protons. A larger dark atom weakens the self-interaction constraints and increases the ionization fraction $f_i$. This latter effect dominates, and electron scattering experiments mostly probe dark electron scattering.

Let's see this explicitly. The ionization fraction $f_i \propto \alpha_d^{-4}$ increases for small dark couplings. However, the coupling is bounded from below by astrophysics and cosmology, see \cref{sec:aDM_cosmo}.
For small mass ratios $R$ and $m_{H'} \lesssim 6$ GeV, the Bullet Cluster bound dominates and gives (with $R \lesssim R_{\rm crit}$ from \cref{Rcrit})
\be
\alpha_d \gtrsim 0.2\, f(R) \(\frac{{\rm GeV}}{m_{H'}}\)^{3/2} \quad \Rightarrow \quad
\alpha_d f_i \lesssim \frac{4\times 10^{-9}}{R\, f(R)^3}  \(\frac{m_{H'}}{{\rm GeV}}\)^{13/2}.
\label{bullet_mH}
\ee
For $R\gtrsim 1$ and $\alpha_d =1$, this indicates the start of the $R=1.1$-line (indicated by the dot) at $m_\chi = m_{H'}/2 = 0.4\,$GeV in the top left panel of \cref{fig:dd-atomic-dm}. The corresponding cross section is
\begin{align}
    \bar \sigma_{\chi e,\mathrm{eff}} 
   & \lesssim 10^{-41} \mathrm{cm}^2 \left(\frac{\epsilon}{10^{-8}}\right)^2 \(\frac{4^3}{Rf(R)^3}\)  \(\frac{m_{H'}}{{\rm GeV}}\)^{13/2}  \(\frac{m_\chi}{m_{H'}}\) \,,
   \label{se_max}
\end{align}
where we used $\xi=0.3$.
For $R\gtrsim 1$, it is about three orders below the current experimental sensitivity.

The maximum effective cross section \cref{eq:sig_eeff} is reached for the highest dark constituent mass that can be probed by experiment, $m_{e'} \sim 1\,$GeV. In this mass region, the dark coupling is set by saturating the ionization constraint $f_i = 0.01$, which  sets a lower bound on the dark gauge coupling
\be
\alpha_d \gtrsim 0.1 \(\frac{m_{H'}}{10^3 \,{\rm GeV}}\)^{1/2} \(\frac{10}{R}\)^{1/4}. 
\label{alpha_low}
\ee
As a result, the maximum effective cross section scales with $R^{-3/4}$ for dark electrons. For dark protons, increasing $R$ shifts the constituent mass outside the experimental sensitivity of electron scattering experiments and dark electron scattering dominates the phenomenology. For $R=10$, we find the maximum cross section $\bar \sigma_{e' e,{\rm eff}} \lesssim 3 \times 10^{-34}\mathrm{cm}^2 $ around $m_{e'} \approx m_{H'}/R \approx 1\,$GeV. 

For small dark electron masses, it becomes increasingly hard to satisfy the Bullet Cluster constraint on the dark coupling \cref{bullet_mH}. Capping the mass ratio at $R \lesssim 4\times 10^3$, there is no available parameter space for masses $m_{e'} \lesssim 0.1 \,$GeV. 

\paragraph{Dark fermion-nucleon scattering}

The effective dark fermion-nucleon cross section, based on \cref{ref_proton_scat0}, is 
\begin{align}\label{eq:df-n-eff}
 \bar \sigma_{\chi n,\mathrm{eff}} \, m_d^4 
   & = 2.4\times 10^{-34} \mathrm{cm}^2\, \mathrm{MeV}^4 \(\frac{\alpha_d}{0.1}\) \(\frac{\epsilon}{10^{-8}}\)^2  \(\frac{\mu_{\chi p}}{m_p}\)^2 
   \(\frac{m_\chi}{m_{H'}}\)f_i(m_H',R,\alpha_d)\,.
\end{align}
Just as for electron scattering, the cross sections and rates for dark electron and dark proton scattering off nucleons are equal up to the small difference in the reduced mass. The dark proton mass is bounded from below by astrophysical constraints; for sub-GeV constituent masses only dark electron scattering off protons is viable. For larger masses, dark proton scattering dominates the phenomenology, thanks to the $m_\chi/m_{H'}$-factor that corrects for the abundance.

Since $\bar \sigma_{p'n,\mathrm{eff}}\propto \alpha_d f_i$, the largest effective cross section is obtained for the largest dark gauge coupling that satisfies \cref{alpha_low}, i.e. for large constituent mass and small $R$. We have chosen these parameters in \cref{eq:df-n-eff} and read off that the effective dark proton cross section is limited to $\bar \sigma_{p'n,\mathrm{eff}}\,m_d^4 \lesssim  2.4 \times 10^{-36} \mathrm{cm}^2\, \mathrm{MeV}^4$. 
This upper bound only depends weakly on $R$ through \cref{alpha_low}.
The effective dark electron cross section for the same constituent mass is a factor $R^{-1/2}$ smaller than for dark proton scattering. This happens due to a combination of two changes: the dark atom mass in \cref{alpha_low} is a factor of $R$ larger, which introduces a relative factor $R^{1/2}$; and there is an explicit factor $m_{e'}/m_{H'}\propto R^{-1}$ in \cref{eq:df-n-eff}.
The above analysis agrees with the numerical results in the top-right panel of \cref{fig:dd-atomic-dm}.

\paragraph{Dark atom-electron scattering}
The elastic dark atom-electron effective cross section \cref{eq:sigma_Hf_el} for $R\gg  1$ is
\begin{align}
    \sigma^{\rm el}_{H'e} &= 9.3\times 10^{-48} \mathrm{cm}^2 \, \left(\frac{R}{10}\right)^4 \left(\frac{1}{\alpha_d}\right)^3 \left(\frac{\epsilon}{10^{-8}}\right)^2 \left(\frac{\GeV}{m_{H'}}\right)^4 \left(\frac{\mu_{H' e}}{m_e}\right)^2 .\label{eq:xsect-He-numeric}
\end{align}
The rate is proportional to $a_b^4 \sim (\alpha_d m_{H'}/R)^{-4}$, which explains the scaling: the smaller the dark gauge coupling and/or larger $R$, the tighter the atom is bound, and the weaker the cross section. For sub-GeV scale masses the Bullet Cluster constraint can only be satisfied for perturbative couplings in the limit $m_{H'}\sim {\rm GeV}$ and $R \sim 1$; the corresponding dark atom-electron cross section is out of reach of future detectors.

\paragraph{Dark atom-nucleon scattering}

The elastic dark atom-nucleon effective cross section  \cref{eq:sigma_Hf_el} for $R\gg 1$ is
\begin{align}
    \sigma^{\rm el}_{H'n} &= 3.1\times 10^{-38} \mathrm{cm}^2 \, \left(\frac{R}{100}\right)^4 \left(\frac{0.1}{\alpha_d}\right)^3 \left(\frac{\epsilon}{10^{-8}}\right)^2 \left(\frac{10\,\GeV}{m_{H'}}\right)^4 \left(\frac{\mu_{H' p}}{m_p}\right)^2. \label{eq:xsect-Hf-numeric}
\end{align}
To satisfy the Bullet Cluster constraint for perturbative couplings $\alpha_d < 1$, 
the dark atom should be heavy enough. This is reflected in \cref{fig:dd-atomic-dm}, where the dots mark the end of the lines for fixed $R$ determined by perturbativity.
The cross section is maximized for large mass ratios $R$ which increase the Bohr radius; as before we cap at $R=4000$ because self-interaction predictions become unreliable.

For a fixed $R$, at large masses the gauge coupling is bounded from below from the requirement $f_i\lesssim 10^{-2}$ and for large $R \gtrsim 100$ by the reionizaiton constraint, while for smaller masses the self-interactions give the dominant bound. 
The cross section is peaked on the boundary of these mass regions. For larger $R$ the peak shifts to larger masses and peaks at higher values. 

Let's focus on large $R \sim 10^3$. Reionization is avoided if $\alpha_d|_{\rm re-ion} \gtrsim 10^{-3} \sqrt{R}$, while Bullet Cluster constraint (assuming dominated by the s-wave contribution) gives $\alpha_d|_{\rm Bullet} \gtrsim 0.17 R/m_{H'}^{3/2}$. At large $R$ the $p$-wave contribution may become important for Bullet Cluster velocities, and the dominant self-interaction constraint comes instead from small-scale structure; this relaxes the lower bound on the dark gauge coupling by about a factor 10. Here we only try to understand the qualitative behavior, and for simplicity we neglect this (but the effect can be seen in bottom-right panel of \cref{fig:dd-atomic-dm}).
The boundary between the regimes $\alpha_d|_{\rm re-ion}=\alpha_d|_{\rm Bullet}$ is for $m_{H'}\sim 30 R^{1/3}$. This choice minimizes the coupling and maximizes the Bohr radius while satisfying all constraints, and thus maximizes the cross section. 
We take $R = 10^2, \,m_{H'} \sim 50\,$GeV, and $\alpha_d \sim 0.04$, which saturates the Bullet Cluster (and actually, also small scale structure constraints) and ionization constraints at the boundary region; the maximum cross section is
\begin{align}
    \sigma^{\rm el}_{H'n} &= 7.8 \times 10^{-36} \mathrm{cm}^2 \, \left(\frac{R}{10^3}\right)^4 \left(\frac{0.04}{\alpha_d}\right)^3 \left(\frac{\epsilon}{10^{-8}}\right)^2 \left(\frac{50\,\GeV}{m_{H'}}\right)^4.
\end{align}
Overall, nuclear scattering experiments such as XENONnT and PandaX4T have excellent sensitivity to Atomic Dark Matter. Low-threshold experiments such as SENSEI compete with astrophysics and are sensitive mostly to dark electron scattering.

\subsection{Discussion}
\label{SEC:aDM_discussion}

\begin{figure}[t!]
   \centering
   \includegraphics[height=0.47\textwidth]{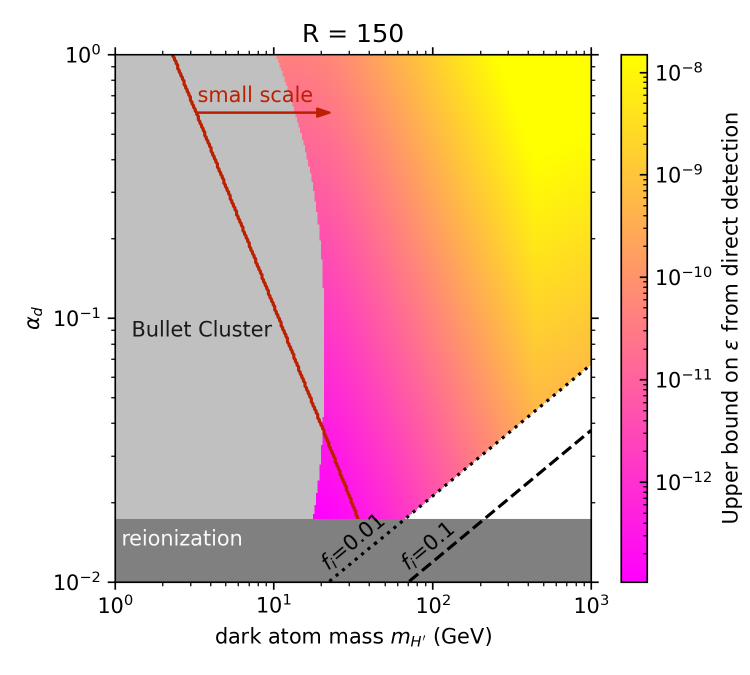}\hfill
   \includegraphics[height=0.47\textwidth]{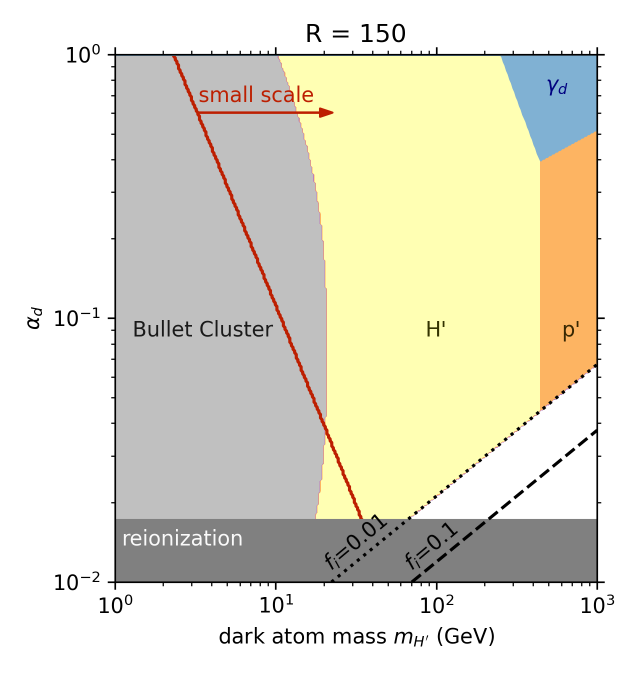}
   \includegraphics[height=0.47\textwidth]{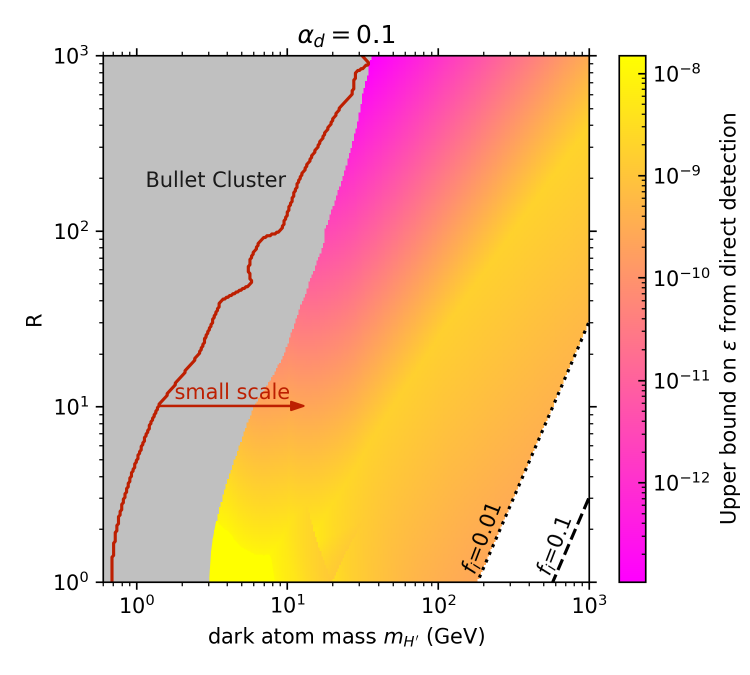}\hfill
   \includegraphics[height=0.47\textwidth]{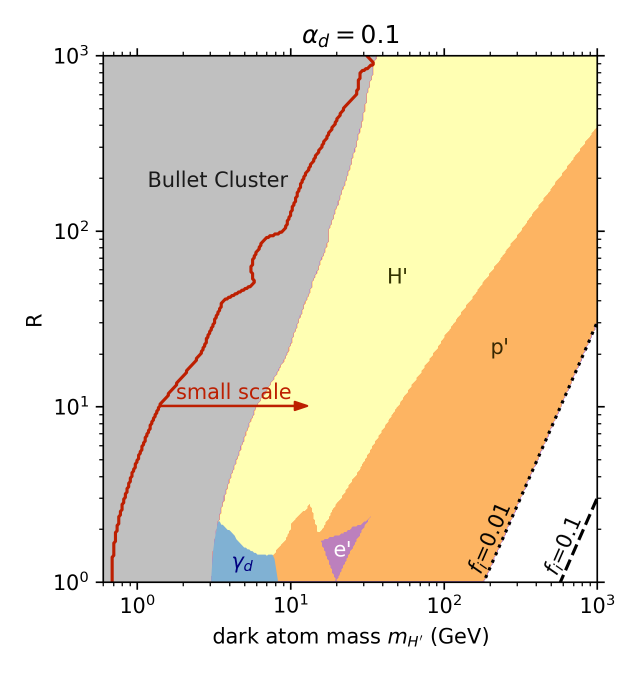}
   \caption{Parameter space of Atomic Dark Matter in the $(m_{H'},\alpha_d)$ plane for a fixed constituent mass ratio $R = 150$ (upper row) and in the $(m_{H'},R)$ plane for a fixed dark coupling $\alpha_d = 0.1$ (lower row). Grey areas are excluded due to effects on the Bullet Cluster and sizeable reionization. Constant values of the ionization fraction $f_i$ today are shown as dotted and dashed black lines. Parameter points to the right of the red line satisfy small-scale structure observations. Left panels: Upper bounds on kinetic mixing $\epsilon$ from current direct detection experiments (colored area).
    Right panels: The dominant direct detection signal is due to scattering of dark atoms (yellow), dark protons (orange), dark electrons (purple), or absorption of solar dark photons (blue).
   \label{fig:result-atom-R150}}
\end{figure}

The interplay of dark atom scattering, constituent scattering and dark photon absorption in a direct detection signal is apparent from \cref{fig:result-atom-R150}. We show the viable parameter space of Atomic Dark Matter for a fixed constituent mass ratio $R=150$ (upper panels) and for a fixed dark coupling $\alpha_d = 0.1$ (lower panels). Constraints from the Bullet Cluster, small-scale structure, dark-atom reionization, and requiring a small fraction of ionized dark atoms today are imposed. For fixed $R=150$, requiring no reionization imposes a lower bound on the dark coupling, $\alpha_d \gtrsim 0.02$, see~\eqref{reion}. Demanding a small relic ionization fraction $f_i \leq 0.01$ excludes regions of small $R$ and large dark atom masses $m_{H'}$, see~\eqref{ion}.
 
For $R = 150$ (upper panels), constraints from the Bullet Cluster and small-scale structure exclude dark atoms with masses below about 10\,GeV, while having relatively little impact on the parameter space of heavier dark atoms. For smaller constituent mass ratios, dark atom masses down to a few GeV are viable (lower panels).

In the left panels of \cref{fig:result-atom-R150}, we illustrate the sensitivity of current direct detection experiments to Atomic Dark Matter. The sensitivity is quantified by the corresponding upper bound on kinetic mixing $\epsilon$ from the most sensitive search for the various scattering and absorption processes, see \cref{sec:direct_detection}. The highest sensitivity to Atomic Dark Matter from scattering is obtained at large mass ratios $R$ and small dark couplings $\alpha_d$, where kinetic mixing is most constrained. For scattering, the dark photon mass plays no role within the relevant parameter range, $10^{-4}\,\text{eV} < m_d < 10^3\,$eV, since the typical momentum transfer is always larger than the mass. For absorption, the displayed bounds on $\epsilon$ are chosen such that absorption does not exceed scattering in sensitivity, which restricts the dark photon mass to $m_d \gtrsim 10^{-4}\,$eV, see \cref{eps_Sun}. For larger dark photon masses, absorption can exceed scattering in sensitivity.

In the right panels of \cref{fig:result-atom-R150}, we show which process dominates the bound on kinetic mixing in the left panels.  For dark atom masses close above the Bullet Cluster bounds, dark atom-nucleon scattering dominates the signal rate. This is due to the highly suppressed abundance of dark constituents, compared to the dark atoms, even though the scattering cross section of the constituents is larger. As $m_{H'}$ or $\alpha_d$ increases, the sensitivity to dark atom scattering decreases because dark constituents are more tightly bound, as predicted by \cref{eq:xsect-Hf-numeric}.
 
For larger dark atom masses, dark proton-nucleon scattering dominates the sensitivity of current direct detection experiments. Also here, the scattering rate decreases as $\alpha_d$ grows, this time due to the reduced ionization fraction~\cref{ion}, see \cref{eq:df-n-eff}. Since the cross sections of dark proton and dark atom scattering scale in the same way with $\alpha_d$, see \cref{eq:df-n-eff,eq:xsect-Hf-numeric} together with \cref{ion}, the relative sensitivity to these processes is independent of $\alpha_d$. This behavior is reflected by the vertical boundaries between $H'$- and $p'$-dominated regions in the upper right panel. Dark electron scattering is suppressed compared to dark proton scattering in most of the parameter space, mostly due to the smaller reduced mass that determines the cross section. Only at small constituent mass ratios $R$ can dark electron scattering dominate the sensitivity to Atomic Dark Matter (see lower right panel). At large dark couplings and small mass ratios $R$, solar dark photon absorption dominates in sensitivity, as the absorption rate increases with $\alpha_d$ and dark atom scattering decreases when $R$ is lowered, see \cref{eq:xsect-Hf-numeric}.

As we have seen, the relative sensitivity of direct detection experiments to Atomic Dark Matter depends on a non-trivial interplay of the various scattering and absorption processes. Moreover, different experimental searches have different sensitivity to the interaction cross section, which introduces another non-trivial dependence. Due to constraints from astrophysics, viable scenarios of Atomic Dark Matter favor mass scales above a few GeV, where dark matter-nucleon scattering dominates and dark matter-electron scattering currently plays no role in probing the parameter space.

Given the current experimental sensitivity, simultaneous dark atom and dark proton scattering off nucleons is possible along the boundaries of the $H'$- and $p'$-dominated regions. In certain regions of parameter space with small $R$ and small dark atom mass, simultaneous scattering of even all three species $H'$, $p'$ and $e'$ can occur (in the lower right panel where the corresponding regions nearly intersect). In this case, direct detection experiments would observe a light-mediator S2-only signal from dark electron scattering, a light-mediator S1-S2 signal from dark proton scattering, and a WIMP-like signal from dark atom scattering, all with similar signal strength.

Simultaneous absorption and scattering can occur anywhere in the yellow, orange and purple regions in \cref{fig:result-atom-R150}. However, it strongly depends on the dark photon mass, which determines the solar flux and thereby the relative sensitivity of absorption and scattering to kinetic mixing, see~\cref{eps_Sun}. Due to the different signals, S2-only for absorption and combined S1-S2 for scattering, both processes can be distinguished in xenon-based experiments. This difference can be used to determine the origin of the signal and lift degeneracies in the large space of dark matter models. In a small region of parameter space with small $R$ and dark proton masses below about 6\,GeV, absorption and scattering are indistinguishable because both signatures are detected through S2-only signals. In this case, an analysis of the energy spectra could help distinguishing the two sources of the signal.




\section{Conclusions}
\label{SEC:conclusions}
In this work, we have investigated the complementarity of dark matter scattering and dark-photon absorption in direct detection experiments for Dirac Dark Matter and Atomic Dark Matter. In both models, bounds from astrophysics - the Bullet Cluster, small-scale structure and (for Atomic Dark Matter) reionization - constrain the expected event rates for dark matter scattering. Dark-photon absorption from the Sun is largely independent from astrophysical constraints, since direct detection searches have become more sensitive to kinetic mixing than stellar cooling.

For Dirac Dark Matter, small-scale structure bounds exclude much of the viable parameter space that could be reached with direct detection searches for dark-matter electron scattering. Sub-GeV Dirac Dark Matter can still be detected through dark-photon absorption, provided that the dark-photon mass lies below the effective mass of the photon in the solar plasma. Dark matter with masses above a few GeV can be probed through nucleon scattering and dark-photon absorption in the near future. This region of Dirac Dark Matter includes the interesting scenario where the relic dark matter abundance is set through freeze-in. Whether scattering or absorption would be observed first does critically depend on the strength of kinetic mixing and on the improvements in detection sensitivity.

For Atomic Dark Matter, the phenomenology at direct detection experiments is much richer, comprising scattering of dark atoms and dark constituents, as well as dark-photon absorption from the Sun. Astrophysics bounds require dark atoms to be heavier than a few GeV for nearly degenerate constituent masses, and even heavier for larger constituent mass ratios. The relic constituents themselves, dark electrons and dark protons, could be observed through electron scattering and nucleon scattering in the near future. This long-range interaction, mediated by a light dark photon, is much stronger than the short-range interaction of dark atoms with nuclei. Atomic Dark Matter could still be first observed through dark atom-nucleon scattering, since dark constituents make up less than a percent of the total dark matter abundance today; otherwise their strong self-interactions would induce too large effects in small-scale structure. Dark-photon absorption is observable in most of the parameter space of Atomic Dark Matter, as long as kinetic mixing is strong enough. Remarkably, in parts of the parameter space simultaneous signals of all four particles are possible.

Our predictions contain two main messages for future dark matter searches at direct detection experiments: First, dark matter scattering through light vector mediators in the meV-to-keV range entails the possibility of mediator absorption. If the strength of both signals is comparable, they can be disentangled by either using spectral information in S2-only analyses, or by adding a second detection mode for scattering, as in S1+S2 analyses. Second, observing an absorption signal not only indicates the mass of the dark photon, but also gives access to its interaction strength, provided that the particle is produced in the Sun. If at the same time a scattering signal is seen, the combination of both observations suggests that the absorbed dark photon is indeed the mediator between dark matter and the Standard Model.

In this way, combined interpretations of direct detection data can achieve an enhanced sensitivity to dark sectors, compared to analyzing individual searches independently of one another. We look forward to what this strategy might reveal in the near future.

\acknowledgments
We thank Wim Beenakker for sharing his wisdom about atom scattering with us. The work of PB and MP was funded by NWO-klein2 grant OCENW.KLEIN.427.

\clearpage

\appendix
\section{Dark atom-proton scattering}\label{sec:app1}

In this appendix, we sketch our calculation of the cross section for elastic $H' + p \to H' + p$ scattering, where $H'$ is dark atom and $p$ is the SM proton. The dark atom is composed of a dark electron $e'$ and a dark proton $p'$. The calculation for dark atom-electron scattering is analogous and can be obtained by replacing $p \to e$. Here, we utilize the feebleness of $\epsilon$ to decouple the scattering dynamics, governed by $\epsilon$, from the internal dynamics of dark atoms, governed by $\alpha_d$. This approach has the advantage compared to the Born-Oppenheimer approximation, that it allows us to compute the cross section even in the degenerate mass limit for the constituents $R \to 1$, extending the results presented in \cite{Cline:2012is}.

There are two possible scattering processes: elastic and inelastic. Due to the low momenta involved, for all parameter points we consider, the collisional kinetic energy of the incoming dark atom is smaller than its excitation energy, specifically
\begin{align}
    \frac12 m_{H'} v^2 < \frac34 B_{H'} = \frac38 \alpha_d^2 \mu_{H'}, \label{eq:no-excitation}
\end{align}
where $v \sim 10^{-3}$ is the relative collision velocity and $B_{H'}$ is the dark atom binding energy. Consequently, ground state ($n=1$) dark atoms cannot be excited to the $n=2$ state, and inelastic scattering only occurs via flipping the spins of the dark constituents. This, however, is only relevant for $R$ very close to 1 where the elastic scattering cross section vanishes due to the cancellation between the dark constituents charge distributions.

The non-relativistic Hamiltonian for the $H'-p$ system is
\begin{align}
     H = \sum_{i={e',p',p}} \frac{\bm{p}^2_i}{2 m_i} + V_d (|\bm r_{e'} - \bm r_{p'} |) \pm V_\epsilon (|\bm r_{p} - \bm r_{p'}|) \mp V_\epsilon (|\bm r_{p} - \bm r_{e'}|).
\end{align}
Here, the sign of $V_\epsilon$ depends on the charges of the involved particles, and
\begin{align}
    V_d (r) = - \frac{\alpha_d}{r} e^{-m_d r},
    \quad
    V_\epsilon (r) = \frac{\epsilon \sqrt{\alpha\alpha_d}}{r} e^{-m_d r} \equiv \frac{\alpha_\epsilon}{r} e^{-m_d r}.
\end{align}
It is convenient to use the following coordinates,
\begin{align}
    \bm \R = \frac1M (m_{e'} \bm r_{e'} + m_{p'} \bm r_{p'} + m_{p} \bm r_{p}),
    \quad
    \bm r = \bm r_{e'} - \bm r_{p'},
    \quad
    \bm \R_p = \bm r_p - \bm \R,
\end{align}
with $M = m_{e'} + m_{p'} + m_p$ and the positions of the various fermions, $\bm r_{i}$. They describe the center-of-mass coordinates, $\bm \R$, the relative distance between the dark electron and dark proton, $\bm r$, and the distance of the center of mass to the scattering particle $p$, $\bm\R_p$. The corresponding momenta are $\bm\P, \bm p,$ and $ \bm \P_p.$
With these new coordinates, the Hamiltonian becomes
\begin{align}
     H = H_{\bm\R} + H_{\bm r} + H_{\bm \R_p,\bm r},
\end{align}
where
\begin{align}
    H_{\bm\R} &= \frac{\bm \P^2}{2M},\nn\\
    H_{\bm r} &= \frac{\bm p^2}{2\mu_{H'}} + V_d(r),\nn\\
    H_{\bm \R_p,\bm r} &= \frac{\bm \P_p^2}{2\tilde\mu_{H'p}}   \pm V_\epsilon \left(\left|\bm \R_{p} + \frac{1}{1+R} \bm r\right|\right) \mp V_\epsilon \left(\left|\bm \R_{p} - \frac{R}{1+R}  \bm r \right|\right).
\end{align}
Here, $R = m_{p'}/m_{e'}$ and $\tilde\mu_{H'f} = (m_{e'} + m_{p'}) m_p / (m_{e'} + m_{p'} + m_p).$ The motion of the $H'-p$ system as a whole is described by $H_{\bm \R}$ and does not affect the scattering process.

In the relevant parameter space of Atomic Dark Matter we have $\epsilon \sqrt{\alpha} \ll \sqrt{\alpha_d}$, and we can consider $V_\epsilon$ as a small perturbation to $V_d$. At the lowest order in this perturbation, the total wave function can be factorized as
\begin{align}
    \Psi(\bm\R,\bm r, \bm \R_p) = \varphi(\bm\R) \phi(\bm r) \psi(\bm \R_p),
\end{align}
where $\phi(\bm r)$ is fully described by $H_{\bm r}$. In our parameter space of interest, $a_b \ll 1/m_d$, where $a_b = (\alpha_d \mu_{H'})^{-1}$ is the Bohr radius of the dark atom, and $H_{\bm r}$ becomes the Hamiltonian for a hydrogen-like atom. In its ground state, the corresponding wavefunction is
\begin{align}
    \phi_{100} (r) = \frac1{\sqrt{\pi a_b^3}} e^{-r/a_b}.
\end{align}
Multiplying both sides of the Schr\"odinger equation $H \Psi = E_{\rm{tot}} \Psi$ by $\varphi^*(\bm\R) \phi^*(\bm r)$ and integrating over $\bm \R$ and $\bm r$, we obtain an equation for the scattering component $\psi(\bm \R_p)$,
\begin{align}
    \left[\frac{\bm \P_p^2}{2\tilde\mu_{H'p}} + V_p(\bm \R_p) \right] \psi(\bm \R_p) = E\, \psi (\bm \R_p)\,,
\end{align}
where the scattering potential is
\begin{align}
    V_p(\bm \R_p) = \pm \int \mathrm{d} \bm r^3 \phi^*(\bm r) \left[V_\epsilon \left(\left|\bm \R_{p} + \frac{1}{1+R} \bm r\right|\right) - V_\epsilon \left(\left|\bm \R_{p} - \frac{R}{1+R}  \bm r \right|\right) \right] \phi(\bm r).
\end{align}
In the case of scattering with ground-state dark atoms and $a_b \ll 1/m_d$,
\begin{align}
    V_p (\R_p) = \mp\frac{\alpha_\epsilon}{a_b} \left[ \left(\frac{a_b}{\R_p} + 1 + R \right) e^{-2\left(1 + R\right)\R_p/a_b} - \left(R \to \frac1R \right)\right]. \label{eq:pot-atom-SM}
\end{align}
In the Born approximation
\begin{align}
    \alpha_\epsilon\, \mu_{H' p}\, a_b \,\frac{R-1}{R+1} \ll 1\,,
\end{align}
one can compute the elastic $H'-p$ scattering cross section from the potential \eqref{eq:pot-atom-SM} and obtain the elastic scattering cross section $\sigma_{H' p}^{\rm el}$ in \cref{eq:sigma_Hf_el}. In the limit $m_{p'} \gg m_{e'}$, or $R \gg 1$, we reproduce the potential and cross section presented in Ref.~\cite{Cline:2012is}. In the limit $R \to 1,$ the potential \eqref{eq:pot-atom-SM} vanishes and inelastic scattering dominates.

\bibliographystyle{JHEP_improved}
\bibliography{main}

\end{document}